\documentclass[letterpaper,12pt,oneside,onecolumn]{article}
\usepackage{geometry}%
\usepackage{amssymb}
\usepackage[figuresright]{rotating}

\begin{document}%

\newcommand{\spd}{$sp$--$d$ }
\newcommand{\ef}{$E_{\rm F}$ }

%
%
%
\begin{center}

{\bf \Large Electronic structure of complex $spd$ Hume-Rothery phases
in transition-metal aluminides}

\vspace{1cm}
{\large Guy Trambly de Laissardi\`ere$^1$,
Duc Nguyen-Manh$^2$, Didier Mayou$^3$}%

\vspace{0.7cm}
{$^1$
Laboratoire de Physique Th\'eorique et Mod\'elisation,
CNRS / Universit\'e de Cergy--Pontoise, 95302 Cergy--Pontoise\\

$^2$ UKAEA Culham Division, Culham Science Centre,
Abingdon,
OX14 3DB,\\ United Kingdom \\

$^3$ Laboratoire d'Etudes des Propri\'et\'es Electronique des Solides,
CNRS, B.P. 166, 38042 Grenoble Cedex 9, France}
\end{center}


\hspace{1.8cm} electronic structure%

\vspace{0.5cm}

The discovery of quasicrystals phases and approximants in
Al(rich)--Mn system has revived the interest for complex aluminides
containing transition-metal atoms. On one hand, it is now
accepted that the Hume-Rothery stabilization plays a crucial role.
On the other hand, TM atoms have also a very important effect on
their stability and their physical properties.
In this paper, we review studies that unifies
the classical Hume-Rothery stabilization for $sp$ electron phases
with the virtual bound state model for transition-metal atoms
embedded in the aluminum matrix.
These studies lead to a new theory for ``$spd$
electron phases''.
It is applied successfully to
Al(Si)--transition-metal alloys and it gives a coherent picture of
their stability and physical properties.
These works are based on
first-principles calculations of the electronic structure and
simplified 
models,
compared to experimental results.

A more detailed review article is published in Prog. Mater. Sci. 
50 (2005) p. 679-788.

\tableofcontents

\section{Introduction}

Aluminides can form a large variety of atomic complex structures
among them transition metal aluminides are of
particular interest: they have potential applications due to their
high strength and light weight.
Some of these phases are
quasicrystalline, thermodynamically stable, and present a very
high structural quality and specific properties.
At the same time their phase diagrams are
complex and in particular that of Al--Mn is still uncomplete. These
are often structurally complex and bear resemblance with
medium-range structure of quasicrystalline phases. All these facts
raise fundamental questions concerning the electronic structure
and the stability of transition-metal aluminides and also the
occurrence of quasiperiodicity among these phases.

Very early, G. V. Raynor \cite{Raynor49a} has proposed to extend the
Hume-Rothery concept
\cite{Hume-Rothery54,Massalski78,Paxton97}
to $sp$ alloys containing transition-metal
atoms (TM atoms,
$\rm TM=Ti$, V, Cr, Mn, Fe, Co, and Ni).
He showed that the rules for the appearance of
crystalline structures as a function of an average number of
valence electrons per atom $e/a$ could be extended provided that a
\textit{``negative valence''} was given to transition metal atoms.
He interpreted this as a negative charge
that was located
on the transition metal atom and a filled $d$ band. This
interpretation is not appropriate since it would correspond to a
great electrostatic energy and indeed experiment show that it is
not the case. Another difficulty with a too naive extension of the
standard theory of Hume-Rothery phases concerns the experimental
values of the density of states. The density presents a minimum at
the Fermi energy (\textit{``pseudogap''}) in most systems. Yet one
expects that in a nearly-free electron gas the $d$ orbitals of
transition metal give a strong contribution to the total density
of states in accordance with the Virtual Bound States model of
the impurity limit
(J. Friedel \cite{Friedel56} and P.W. Anderson \cite{Anderson61}).
In this model there is only a very small
transfer at the opposite of Raynor assumption. Thus, it appears
that the role of transition metal atoms in aluminides is explained
neither by the Raynor assumption, nor by the standard Virtual
Bound States model.
Now the Al(rich)--TM quasicrystals and related phases are
often considered as Hume-Rothery alloys
(see for instance
\cite{Friedel87,Fujiwara89,Fujiwara91,Tsai05_TMS,Mizutani05_TMS}),
but the problem
of the physical interpretation of
electronic structure of TM atoms still persists.

In this paper we summarize our work on the electronic
structure of Al(rich)--TM phases that lead
to consider these aluminides as
\textit{``$spd$ Hume-Rothery compounds''}.
Our study is based on numerous ab initio calculation
on band structure calculations
(using the  self-consistent tight-binding
Linear Muffin-Tin Orbital (LMTO)
method~\cite{Andersen75}), and comparison with simple
but physical Hamiltonian models.
A more detailed review article is in
preparation \cite{PMS}.

\section{Density of states of $spd$ electron phases}

In the literature, there are a lot of theoretical studies of
the electronic structure of Al(rich)--TM crystals and Al(rich)--TM
crystalline approximants of quasicrystals
from first-principles.
In this section we focus on properties that are common to those
phases in spite of different atomic structures.
Some of these phases are rather \textit{``simple'' phases} that contain a
small number of atoms in a unit cell, whereas other phases are
\textit{``complex'' phases} that contain large number of atoms
in a unit cell (table \ref{Table_AlMn_LMTO}).

\begin{figure}[t!]
\begin{center}
\begin{tabular}{ccc}

\rotatebox[origin=c]{90}{DOS (states / (eV unit cell))}

&
\begin{minipage}[c]{7.3cm}

\begin{flushright}
\includegraphics[height=4.4cm]{c_al6mn_t_TMS.eps}

\vspace{0.1cm}
\includegraphics[height=4.4cm]{Sugiyama_lm5_all_TMS.eps}
\end{flushright}
\end{minipage}
\vspace{0.1cm}

&
\begin{minipage}[c]{7.3cm}

\begin{flushright}
\includegraphics[height=4.4cm]{c_al7cu2fe_t_TMS.eps}

\vspace{0.1cm}
\includegraphics[height=4.4cm]{a_AlCuFe16_all_TMS.eps}
\end{flushright}
\end{minipage}
\vspace{0.1cm}

\\

 & ~~$E-E_{\rm F}$ (eV) & ~~$E-E_{\rm F}$ (eV)\\
\end{tabular}

\caption{LMTO total DOS orthorhombic of o-$\rm Al_6Mn$,
tetragonal $\omega$-$\rm Al_7Cu_2Fe$,
cubic $\alpha$-Al--Mn and cubic $\alpha$'-Al--Cu--Fe \cite{PMS}.}
\label{LMTO_totalDOS}
\end{center}
\end{figure}

\vspace{0.25cm}
\begin{figure}[!t]
\begin{center}
\hspace{1cm}\includegraphics[width=10cm]{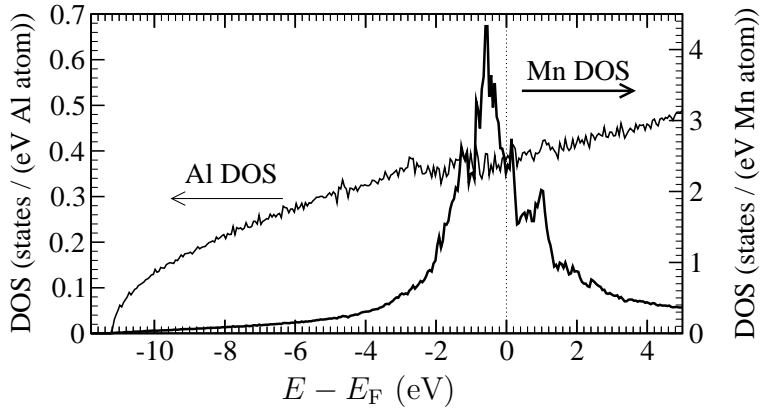}


~~~~$E-E_{\rm F}$ (eV)

\caption{Non-magnetic DOS in a $\rm Al_{107}Mn$ model.
Mn atoms are in
substitution to Al atoms in an  Al f.c.c.
Al(1) and Mn atoms are first-neighbors.}
\label{FigDOS23_Al107Mn}
\end{center}
\end{figure}

\begin{table}[t!]
\begin{center}
\caption{Examples of Al--Mn crystals.}

{$\mathcal{N}$ is the number of atoms per unit cell.
Calculated DOS at the energy $E_{min}$
for which the total DOS reaches its
minimum in the pseudogap:
$n$, total DOS (states\,$/$\,eV atom);
$n_{Al}$, local Al DOS (states\,$/$\,eV Al atom);
$n_{Mn}$, local Mn DOS (states\,$/$\,eV Mn atom).
References of the crystallographic structures are given
in Ref.
\cite{PMS} \label{Table_AlMn_LMTO}}
\begin{tabular}{|l|c|c|c|c|c|c|c|}
\hline
Phase  & Space group & $\mathcal{N}$ & $\%$ Mn  &
\multicolumn{4}{c|}{LMTO DOS at $E_{min}$}   \\
\cline{5-8}
 & & & atoms & $n$ & $n_{\rm Al}$ &  $n_{\rm Mn}$ & Refs. \\
\hline
Al & cubic f.c.c.     &   1 & 0   & 0.30 & 0.30 & -- &
\cite{GuyPRB95} \\
\hline
$\rm Al_{107}Mn$ ${\dag}$ & cubic  & 108 & 0.9 & 0.36 & 0.34 & 2.60
& \cite{PMS} \\
\hline
$\rm Al_{31}Mn$ ${\dag}$  & cubic  & 32  & 3.1 & 0.37 & 0.33 & 1.78
& \cite{PMS}\\
\hline
$\rm Al_{12}Mn$  & cubic & 13 & 7.7 & 0.22  & 0.18 & 0.69 &
\cite{GuyPRB95} \\
 & Im3   & &  & & & &\\
\hline
o-$\rm Al_6Mn$   & orthorhombic  & 14 & 14.3& 0.20 & 0.11 & 0.75 &
\cite{GuyPRB95} \\
 & Cmcm & &  & & & & \\
\hline
$\rm Al_{65.9}Pd_{12.2}Mn_{14.6}Si_{7.3}$ & Cubic & 123 & 14.6 & 0.12
& 0.06 & 0.45 & \cite{Hippert99_JPCM} \\
 & Pm3 & &  & & & & \\
\hline
T-$\rm Al_{79.5}Pd_{5.1}Mn_{15.4}$ & orthorhombic & 156 & 15.4 & 0.21
& 0.10 & 0.79 &
\cite{Hippert99_JPCM} \\
 & Pnma  & &  & & & & \\
\hline
$\alpha$-$\rm Al_{114}Mn_{24}$ & cubic  & 138 &
17.4 & 0.21 & 0.13 & 0.58 & \cite{Fujiwara89,Fujiwara93} \\
 & $\rm Pm\overline{3}$ & &  & & & & \\
\hline
$\mu$-$\rm Al_{4.12}Mn$   & hexagonal  & 568 & 19.4 &
0.38$^{\,\star}$&0.15$^{\,\star}$&1.32$^{\,\star}$& 
\cite{Duc03,PMS}\\
 & $\rm P6_3/mmc$ & &  & & & &  \\
\hline
$\varphi$-$\rm Al_{10}Mn_3$ & hexagonal  & 26 & 23.1 & 0.16 & 0.07 &
0.47 &
\cite{Guy03}\\
 & $\rm P6_3/mmc$ & &  & & & & \\
\hline
$\rm Al_{13}Mn_4$ ${\ddag}$ & monoclinic  & 51 & 23.6 & 0.22 & 0.10 &
0.60 &  \cite{PMS}\\
 & $\rm C2/m$ & &  & & & & \\
\hline
T-$\rm Al_3Mn$ & orthorhombic  & 156 & 30.8 &  0.29 & 0.10 & 0.71 &
\cite{Hafner98a,PMS} \\
 & Pnma & &  & & & & \\ \hline
\end{tabular}
\end{center}

\vspace{-0.3cm}
~~~~~${\dag}$ Hypothetical model for an Mn impurity in the Al matrix: Mn
substituted Al in

~~~~~~~~the Al f.c.c. lattice

~~~~~${\ddag}$ Structure of $\rm Al_{13}Fe_4$ \cite{Duc95}.

~~~~~$^{\star}$ Preliminary results.
\end{table}

At low energy, the total DOS is nearly-free  electrons
like (figure \ref{LMTO_totalDOS}).
These states are mainly $sp$ states of the Al atoms.
The $d$ states of TM
(TM $=$ Ti, V, Cr, Mn, Fe, Co, Ni)
are observed in the middle of the $sp$
band.
In phases containing Cu atoms, the $d$ peak of Cu is strong and
it is located at an energy lower than that of $d$ peak of TM.

\subsection{Pseudogap}

\begin{figure}[t!]
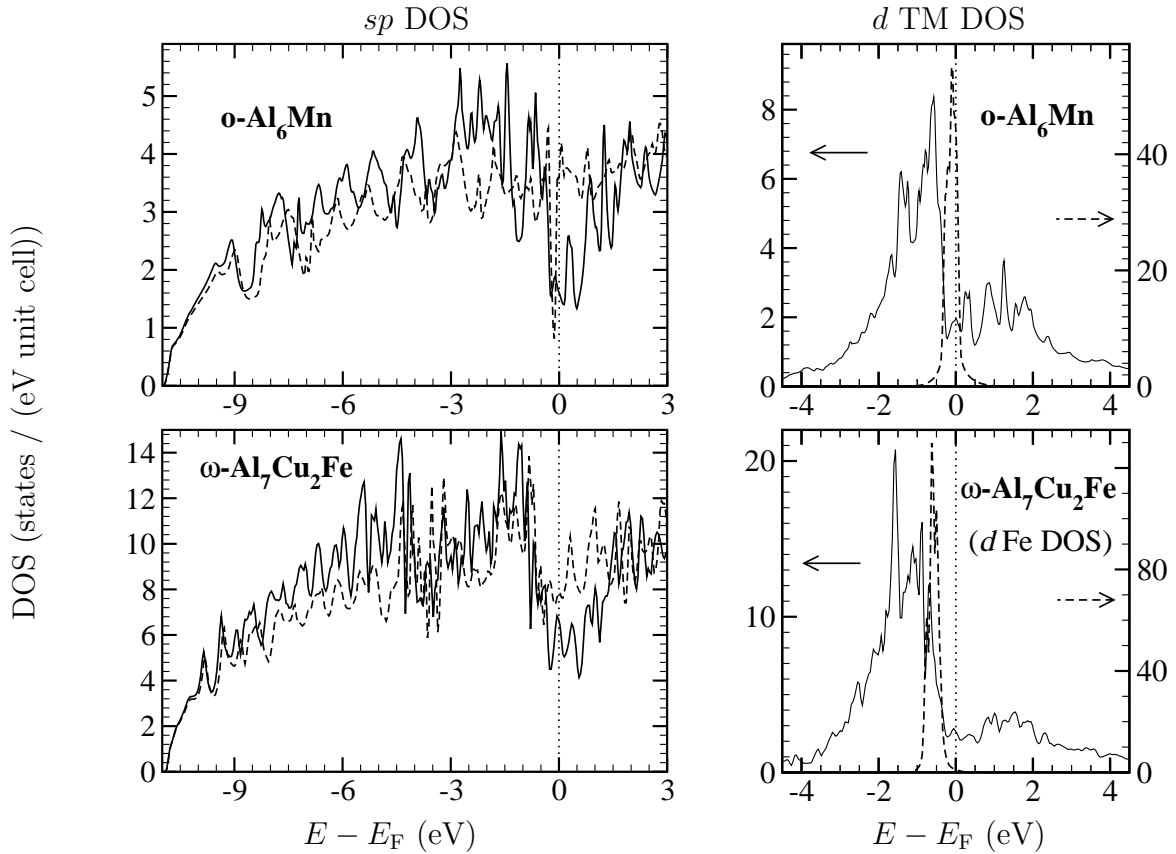

\begin{center}

\begin{tabular}{ccc}
 & ~~~~~~~$sp$ DOS & $d$ TM DOS\\

\rotatebox[origin=c]{90}{DOS (states / (eV unit cell))}

&

\begin{minipage}[c]{8cm}

\begin{flushright}
\includegraphics[height=5cm]{c_al6mn_sp_TMS.eps}

\vspace{0.1cm}
\includegraphics[height=5cm]{c_al7cu2fe_sp_TMS.eps}
\end{flushright}
\end{minipage}
\vspace{0.1cm}

&

\begin{minipage}[c]{6cm}

\begin{flushright}
\includegraphics[height=5cm]{c_al6mn_dmn_TMS.eps}

\vspace{0.1cm}
\includegraphics[height=5cm]{c_al7cu2fe_dfe_TMS.eps}
\end{flushright}
\end{minipage}
\vspace{0.1cm}
\\
 & ~~$E-E_{\rm F}$ (eV) & ~~$E-E_{\rm F}$ (eV)\\
\end{tabular}
\caption{LMTO $sp$ Al DOS and $d$ TM DOS of crystals.
Solid lines, exact calculation;
dashed lines, calculations without $sp$-$d$ hybridization.
\cite{GuyPRB95}}
\label{LMTO_PartialDOS_crystals}
\end{center}
\end{figure}

In many transition-metal aluminides,
the Fermi level
\ef is found near a large valley in the DOS
that splits
the band between bonding states and anti-bonding states
(figure~\ref{LMTO_totalDOS}).
This valley, called \textit{``pseudogap''},
is generally attributed to a combined effect including
the electron diffraction by
the Bragg planes of a prominent Brillouin zone and
a strong
Al($sp$)--TM($d$)
hybridization~\cite{EurophysLet93,GuyPRB95}
(see following section).
As shown in table~\ref{Table_AlMn_LMTO},
this pseudogap is predicted from first-principles in the
Al DOS (mainly $sp$ DOS) of most of the simple phases and
the complex phases (see Refs. in \cite{PMS}).
It is also predicted in the total DOS and the TM DOS (mainly $d$
states) of many phases
(o-$\rm Al_6Mn$, $\alpha$-Al--Mn--Si, \ldots)
(figure \ref{LMTO_PartialDOS_crystals}).
But in the phases containing a concentration of Mn atoms larger
that 20\,\%, the pseudogap can be filled (or partially filled) by
the $d$ states of some TM atoms (T-Al--Pd--Mn, $\rm Al_3Mn$, \ldots)
(figure \ref {LMTO_T_AlPdMn}).
Photo-emission spectroscopy and specific heat measurements
\cite{Belin05_TMS}
have confirmed the presence of pseudogap
in the DOS of many Al--TM quasicrystals and approximants.

\subsection{\spd hybridization}

The $spd$ aluminides are characterized by a strong \spd
hybridization between the Al $sp$ states and the TM $d$
orbitals.
Many experimental studies of photoemission spectra
have shown this property
(E. Belin-Ferr\'e in this proceedings).
It is illustrated from LMTO calculation
by the comparison between the DOSs
calculated with the \spd hybridization
(``exact'' calculation)
and the DOSs calculated by setting to zero the
matrix elements of the Hamiltonian that correspond the
the \spd hybridization
(calculation ``without \spd hybridization'' \cite{Duc92})
(figure \ref{LMTO_PartialDOS_crystals}).
The width of the TM DOS is strongly reduced in the calculation
without \spd hybridization with respect to the exact calculation.
Indeed the width
of the TM DOS (mainly $d$ DOS) is
proportional to the square of the matrix element
that couples the $d$ states and the $sp$ states
in the Hamiltonian

In the case of a TM impurity in the free electron matrix (Virtual
Bound States model) \cite{Friedel56,Anderson61}, the
partial $d$ DOS is a Lorentzian and the free states $sp$
DOS is not modified by the \spd hybridization
(compensation theorem).
This is illustrated by the LMTO calculation
(without spins polarization) for Mn diluted in the Al f.c.c. crystals
(figure \ref{FigDOS23_Al107Mn}).
We have considered a super-cell of Al structure containing 108 atoms.
An Mn substitutes one of the Al atoms in the super-cell, thus as
the concentration is $\rm Al_{107}Mn$.

In Al--TM phases the \spd hybridization is strong too, but
the Virtual Bound States model is no more valid.
Indeed, many theoretical and experimental
studies shows that \spd hybridization plays a crucial role
in the pseudogap formation.
In some cases
($\rm Al_3Ti$, o-$\rm Al_6Mn$, $\omega$-$\rm Al_7Cu_2Fe$,
$\alpha$-Al--Mn, \ldots)
the pseudogap is  present in the calculation without
\spd hybridization and it is increased by the
\spd hybridization (figure \ref{LMTO_PartialDOS_crystals}).
In other cases
($\beta$-$\rm Al_9Mn_3Si$, $\varphi$-$\rm Al_{10}Mn_3$, \ldots)
the pseudogap disappears when \spd hybridization
is suppressed \cite{Guy03}.


Let's remark that in some particular cases the direct $d$-$d$ coupling
between two first-neighbors TM could be important.
But in many Al(rich)--TM alloys
($\rm Al_3Ti$, o-$\rm Al_6Mn$, $\omega$-$\rm Al_7Cu_2Fe$,
$\alpha$-Al--Mn, \ldots)
TM atoms are not first-neighbors, therefore the direct
$d$--$d$ coupling is not enough to explain the generic
properties of the Al--TM DOS.

\subsection{Effect of the TM atom's position}

\begin{figure}[t!]
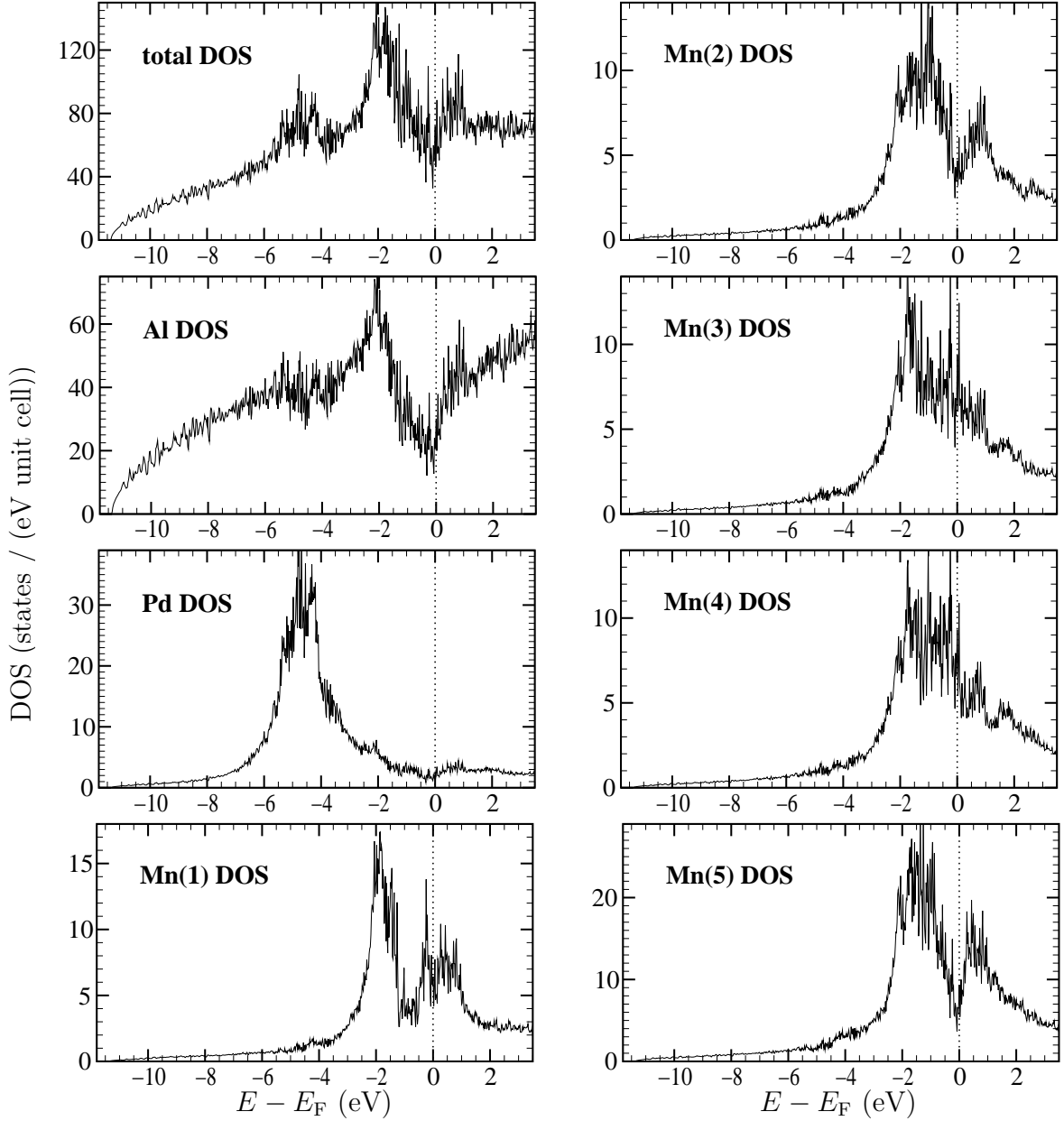

\begin{center}
\begin{tabular}{ccc}

\rotatebox[origin=c]{90}{DOS (states / (eV unit cell))}

&
\begin{minipage}[c]{7.3cm}
\hspace{-0.3cm}
\includegraphics[height=3.9cm]{taylor_tdos.eps}

\vspace{0.1cm}
\includegraphics[height=3.9cm]{taylor_dosAl_TM.eps}

\vspace{0.1cm}
\includegraphics[height=3.9cm]{taylor_dosPd6.eps}

\vspace{0.1cm}
\includegraphics[height=3.9cm]{taylor_dosMn1.eps}
\end{minipage}
&
\begin{minipage}[c]{7.3cm}
\includegraphics[height=3.9cm]{taylor_dosMn2.eps}

\vspace{0.1cm}
\includegraphics[height=3.9cm]{taylor_dosMn3.eps}

\vspace{0.1cm}
\includegraphics[height=3.9cm]{taylor_dosMn4.eps}

\vspace{0.1cm}
\includegraphics[height=3.9cm]{taylor_dosMn5.eps}
\end{minipage}
\\
 & $E-E_{\rm F}$ (eV) & $E-E_{\rm F}$ (eV)\\
\end{tabular}

\caption{LMTO total DOS and local DOSs of the
orthorhombic
T-$\rm Al_{79.5}Pd_{5.1}Mn_{15.4}$
phase, calculated without spin polarization
\cite{Hippert99_JPCM}.}
\label{LMTO_T_AlPdMn}
\end{center}
\end{figure}

In the Virtual Bound State model
\cite{Friedel56,Anderson61},
the $sp$ free states have an uniform amplitude in the real space,
thus the $d$ TM  DOS of  does not depend on the position,
${\bf r}_{TM}$, of TM atom.
But, in Hume-Rothery phases, the diffraction by Bragg planes
is important ($sp$ states are not free states) and the
amplitude of $sp$ states depends on  ${\bf r}$.
Therefore, the effect of the \spd coupling  on the  DOSs depends
on
${\bf r}_{TM}$.
In the literature there are many example of ab initio calculations
showing different local TM DOS of inequivalent TM sites.
In the case of the orthorhombic T-Al--Pd--Mn (figure
\ref{LMTO_T_AlPdMn}), Mn(2) and Mn(5) local DOSs have a pseudogap
near $E_{\rm F}$, whereas Mn(3) and Mn(4) local DOSs do not have.

We have developed a model to calculate the TM DOS in Al--TM
phase which take into account the diffraction of the $sp$ states
by the Bragg planes of a prominent Brillouin zone
\cite{EurophysLet93,GuyPRB95,PMS}.
The TM position is a crucial parameter that can switch on or switch off
the pseudogap in the local  DOS around TM.

\subsection{Effective Mn--Mn interactions induce the pseudogap}

\begin{figure}[t]
\begin{center}
\includegraphics[width=10cm]{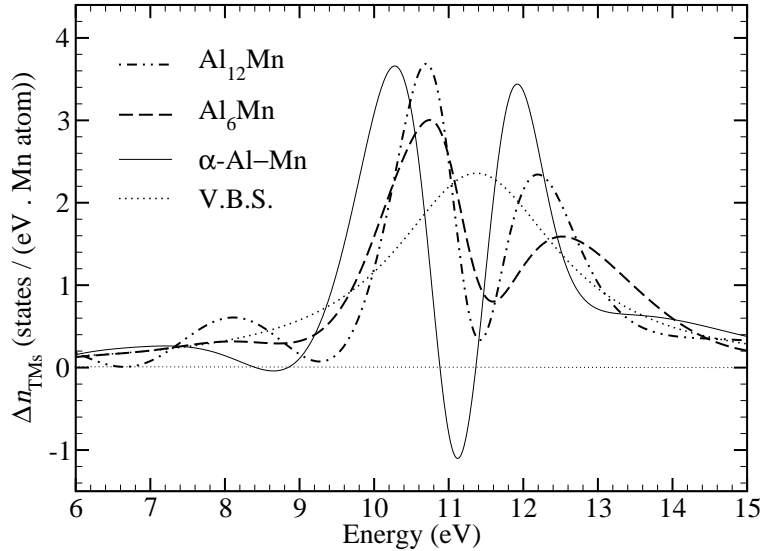}
\caption{Variation of the DOS, $\Delta n_{\rm TMs}$
due to the Mn sub-lattice, in $\rm Al_{12}Mn$, o-$\rm Al_6Mn$,
and $\alpha$-Al--Mn. 
V.B.S. is the case of one Mn impurity in the Al matrix 
(Virtual Bound State). \cite{Guy04_ICQ8}}
\label{DOS_R15_al12mn_al6mn_alpha}
\end{center}
\end{figure}

Considering that the pseudopotential of Al atoms is small,
Al(rich)--TM
phases are modeled by a collection of TM atoms in a jellium (free
electron states of Al). The total DOS, $n$, is written as:
\begin{eqnarray}
n(E) &=&  n_{\rm free}(E) +  \Delta n_{\rm TMs}(E),
\label{DOS_total_Mnpair}
\end{eqnarray}
where, $n_{\rm free}$ is the DOS of free $sp$ states,
and $\Delta n_{\rm TMs}$ the variation of the total DOS due
to the TM atoms.
We calculated
$\Delta n_{\rm TMs}$ as the sum of variation of the DOS
due to each Mn--Mn pair \cite{Guy04_ICQ8}.
When all the TM atoms are on the same Wyckoff position,
$\Delta n_{\rm TMs}$ per TM atoms is:
\begin{eqnarray}
\Delta n_{\rm TMs}(E) &=& \Delta n_{\rm 1TM}(E)
+ \frac{1}{2} \sum_{j\neq1}\Big( \Delta n_{\rm 2TM}(E,r_{1j}) -
2 \Delta n_{\rm 1TM}(E) \Big),
\label{Delta_n_TMs}
\end{eqnarray}
where $j$ is an index the TM atoms.
$\Delta n_{\rm 1TM}$ is the variation of the DOS
due to one TM impurity in free electron 
(Lorentzian, i.e. Virtual Bound State),
and $\Delta n_{\rm 2TM}$, the variation of the
DOS due to two TM atoms in free electrons calculated
by the Lloyd formula (using transfer matrix ${\bf T}$ approach)
\cite{PRB97}.
We have calculated $\Delta n_{\rm Mn}$ for Al--Mn phases
where Mn atoms are not first-neighbors:
$\rm Al_{12}Mn$, o-$\rm Al_6Mn$,
and $\alpha$-Al--Mn.
The sum in (\ref{Delta_n_TMs}) is
computed including the distances $r_{1j}$ up to
the distance $R$.
A well
pronounced  pseudogap appears when $r_{1j}$
up to 10--20\,$\rm \AA$ are taken into account
\cite{Guy04_ICQ8}.
Results with $R=15\,{\rm \AA}$ are presented in
figure \ref{DOS_R15_al12mn_al6mn_alpha}.
For $\alpha$-Al--Mn structure, negative values of
$\Delta n_{\rm Mn}$ are obtained for some energy,
which induces a reduction of the total DOS with
respect to the free electrons DOS, $n_{\rm free}(E)$, at
these energies (equation (\ref{DOS_total_Mnpair})).
These results show that effective medium-range Mn--Mn interaction
contribute to the pseudogap in  those  Al(rich)--TM phases.

\subsection{Negative valence of transition-metal atoms}

In his original
work on negative valence Raynor \cite{Raynor49a} assumed
a transfer of electrons from the
conduction band ($sp$ band) to the $d$ band
in order to compensate the unpaired
spins of the TM elements, and fill the $d$ band.
In this scheme the TM atoms remove electrons from the $sp$
band and thus have a negative valence.
But a  transfer of several electrons on one atom is unrealistic
in metallic alloys since it corresponds to a too large electrostatic
energy for metallic alloys \cite{Hume-Rothery54}.

The LMTO results allow to solve this paradox and
to understand the apparent
negative valence of TM in Al--TM compounds.
Indeed,
there is an increase  of the $sp$ DOS below
\ef  as compared to the free electron DOS
due to the combined effect of
\spd hybridization and the diffraction of $sp$ states
by Bragg planes.
In this scheme
filling these additional $sp$ states below \ef plays the
same role as
filling of the $d$ band in Raynor's scheme.
It results in an apparent negative valence of TM.
Contrary to the $d$ orbitals these additional $sp$ states are
delocalized and do not lead  to a strong electrostatic energy.
Yet one may expect that these additional $sp$ states are
linked to the TM atom and that they follow its displacement. This could
explain the anomalous effective charge of TM elements as
deduced from optical conductivity \cite{OgutPRB96}.

\begin{table}[t!]
\begin{center}
\caption{Negative valence  of transition-metal
elements
in Al(rich) alloys: }

{According to Raynor \cite{Raynor49a}
and quantity ($-\Delta N_{sp}$) calculated from LMTO \cite{EurophysLet93}.}
\label{Tab_ValanceNegLMTO}
\begin{tabular}{|l|c|c|c|c|c|c|}
\hline
 & Cr  & Mn & Fe & Co & Ni \\  \hline
Raynor & $-4.66$ & $-3.66$ & $-2.66$ & $-1.66$ & $-0.66$ \\ \hline
LMTO & $-3.2$  {\small ($\rm Al_{12}Cr$)} &$-2.7$  {\small ($\rm Al_{12}Mn$)} &
$-2.5$ {\small ($\rm Al_7Cu_2Fe$)} &$-1.3$ {\small ($\rm Al_9Co_2$)} &
$-1$ {\small ($\rm Al_3Ni$)}  \\
 & & $-2.0$ {\small ($\rm Al_{6}Mn$)} & &
$-0.9$ {\small ($\rm Al_5Co_2$)} & \\
\hline
\end{tabular}
\end{center}
\end{table}

The variation of the number $d$ of electrons and
$sp$ electrons due to the TM impurity are, respectively,
\begin{eqnarray}
N_{d} = \int^{E_{\rm F}}n_d(\epsilon)\mathrm{d} \epsilon~~{\rm and}~~
\Delta N_{sp} = \int^{E_{\rm F}} \Delta n_{sp}(\epsilon)
\mathrm{d} \epsilon .
\end{eqnarray}
$N_d$ is fixed by the nature of the TM atom.
$\Delta N_{sp}$ is the variation of $sp$ electrons due to
the presence of the TM atoms.
When $sp$ states are free (no diffraction by Bragg planes),
$\Delta N_{sp}=0$, but in actual Al(rich)--TM alloys,
$sp$ states are not free and
$\Delta N_{sp}$ takes a positive value.
From LMTO it can be estimated as (table~\ref{Tab_ValanceNegLMTO}):
\begin{eqnarray}
\Delta N_{sp} = N_{sp} -
N_{sp}({\rm without}~sp \textrm{-}d~{\rm hybridization}),
\end{eqnarray}
The quantity 
\begin{eqnarray}
\mathcal{A}=-\Delta N_{sp}\,,
\end{eqnarray}
given per TM atom, is  what Raynor called
a negative valence of TM.
It corresponds to
an increase of the $sp$ electrons density
around the TM impurity, but it is not
a charge transfer from $sp$ band to $d$ orbitals.
Note that this apparent negative valence depends
on the nature of TM element but also on the Al--TM compounds.
Indeed as seen in the first section
the local TM DOS  depends on the position of the TM atom in the
crystals.

The origin of the additional $sp$ electrons is understandable from a simple
argument based on a sum rule. Consider first the limiting case for which the
diffraction by Bragg plans creates a gap in the DOS at
$E_{\rm F}$.
Consider also $5$ degenerated $d$ orbitals of a non-magnetic
TM impurity in the the free electron matrix (jellium).
When $E_d \ll E_{\rm F}$, it is obvious that
\begin{eqnarray}
N_{d} + \Delta N_{sp} = 10~{\rm electrons},
\label{EqValenceGap}
\end{eqnarray}
since the $d$ band is filled and $\Delta N_{sp}=0$.
If $E_d$ is shifted continuously up a realistic value close to
\ef the eigen energies should be shifting
continuously too. Thus, no states could jump the gap and the total
number of states below \ef is independent of the value of
$E_d$ (when $E_d \leqslant E_{\rm F}$).
Therefore, if there is a gap a \ef, the equation
(\ref{EqValenceGap}) is always satisfied
for $E_d \leqslant E_{\rm F}$,
and one obtains that
\begin{eqnarray}
\mathcal{A}=-(10 - N_{d}) < 0.
\label{EqValenceGap2}
\end{eqnarray}
In actual alloys where there is a pseudogap at $E_{\rm F}$ (not a
gap), one could conjecture that the equality in (\ref{EqValenceGap2})
becomes an inequality:
\begin{eqnarray}
-( 10 - N_{d})< \mathcal{A}< 0,
\label{EqValenceGap3}
\end{eqnarray}
with an apparent valence of TM still negative.

\section{Generalization of the Jones theory for the $spd$ electron phases}

For $sp$ Hume-Rothery alloys, the valence states ($sp$ states)
are nearly-free states scattered by a weak Bragg potential,$V_B$
(Jones theory, see Refs. \cite{Massalski78,Paxton97}).
But, the treatment of Al(rich) alloys containing
TM atoms requires a different model because the
$d$ states of TM are not nearly-free states
\cite{EurophysLet93,GuyPRB95}.
In this section, we show briefly that in $spd$ Hume-Rothery
alloys, the $sp$ electrons feel an
{\it ``effective Bragg potential''}
\cite{GuyPRB95}
that takes into account the strong effect TM atoms
via the \spd hybridization.

Following a classical approximation
\cite{Friedel56,Anderson61}
for Al(Si)-Mn alloys, a simplified
model is considered where $sp$ states are nearly-free
and $d$ states are localized on Mn sites $i$.
The effective Hamiltonian
for the sp states is written:
\begin{eqnarray}
H_{eff(sp)}= \frac{\hbar^2\,k^2}{2m} + V_{B,eff}
\label{Hamil_eff_sp}
\end{eqnarray}
where $V_{B,eff}$ is an effective Bragg potential
that takes into account the scattering
of  $sp$ states by the
strong potential of Mn atoms.
$V_{B,eff}$ depends thus  on
the positions ${\bf r}_i$ of Mn atoms.
Assuming that all Mn atoms are equivalent and that two Mn
atoms are not first-neighbor, one obtains:
\begin{eqnarray}
V_{B,eff}({\bf r}) &=&  \sum_{\bf K}
V_{B,eff}({\bf K}) e^{i{\bf K}.{\bf r}},  \\
V_{B,eff}({\bf K})  &=&
V_B({\bf K}) +
\frac{|t_{{\bf K}}|^2}{E - E_d}
\sum_i e^{-i {\bf K}.{\bf r}_{i}},
\label{EqVeffectif}
\end{eqnarray}
where the vectors ${\bf K}$ belong to the
reciprocal lattice, $t_{{\bf K}}$ is a average
matrix element that
couples sp states ${\bf k}$ and ${\bf k}-{\bf K}$ via the
\spd hybridization, and $E_d$ is the energy of $d$ states.
The term $V_B({\bf K})$ is a weak potential independent of the
energy $E$. It corresponds to the  Bragg potential for
$sp$ Hume-Rothery compounds.

The last term in equation (\ref{EqVeffectif}), is due to
the $d$-resonance of the wave function by the potential of
Mn atoms. It is strong in
an energy range
$E_d-\Gamma \leq E \leq E_d+\Gamma$,
where
$2\Gamma$ is the width of the $d$ DOS.
This term is essential as
it does represent
the diffraction of the $sp$ electrons by a network
of $d$ orbitals,
i.e. the  factor
$\left(\sum_i e^{-i {\bf K}.{\bf r}_{i}}\right)$
corresponding to the structure factor of the
TM atoms sub-lattice.
As the $d$ band of Mn is almost half filled,
$E_F \simeq E_d$, this factor is important
for energy close to $E_F$.
Note that
the Bragg planes associated with the second term of
equation (\ref{EqVeffectif}) correspond to
Bragg planes determined by diffraction.

This qualitative
analysis suggests that both \spd hybridization and
diffraction
of $sp$ states by the sub-lattice of Mn atoms are essential
to understand the electronic structure of
Al(Si)-Mn alloys.
The strong effect of \spd hybridization
on the pseudogap is then  understood
in the framework of  Hume-Rothery mechanism.

\section{Stability of complex $spd$ electron phases}

\subsection{Ab initio studies of the phase stability}

First-principles electronic structure calculations have
proven to be an accurate and efficient tool to understand
the physics of materials in particular in investigating
systematically the
energetics and (meta)stability of transition-metal aluminides
alloys~\cite{Nguyen95,Duc97,Duc99a,Duc99b}.
Although there have been several
first-principles calculations for different
Al(rich)--TM compounds and related
quasicrystalline phases, it is
desirable to elucidate why
the quasicrystalline phase is stable only by forming with
$TM$ of group VIIA (Mn, Re)
and group VIIIA (Fe, Ru, Os, Co, Rh, Ir, Ni)?
It is also essential to know whether these calculations
confirm (or not) a Hume-Rothery mechanism for stabilizing $spd$
compounds
like that has been shown for simple
$sp$ compounds.

In order to analyze the trend in structural stability
of the transition-metal trialuminides  across the transition
metal series,
the relative stability between the simple
tetragonal I4$/$mmm  $\rm Al_{3}$$TM$
($\rm Al_3V$ structure)
and the complex  monoclinic C2$/$m
$\rm Al_{13}$$TM_{\rm 4}$
($\rm Al_{13}Fe_4$ structure)
has been investigated
using first-principles
total energy and electronic structure calculations
within the LMTO method~\cite{Duc95}.
$\rm Al_{3}$$TM$ has a similar composition than
$\rm Al_{13}$$TM_{\rm 4}$,
but a  simpler tetragonal structure.

\begin{figure}[t!]
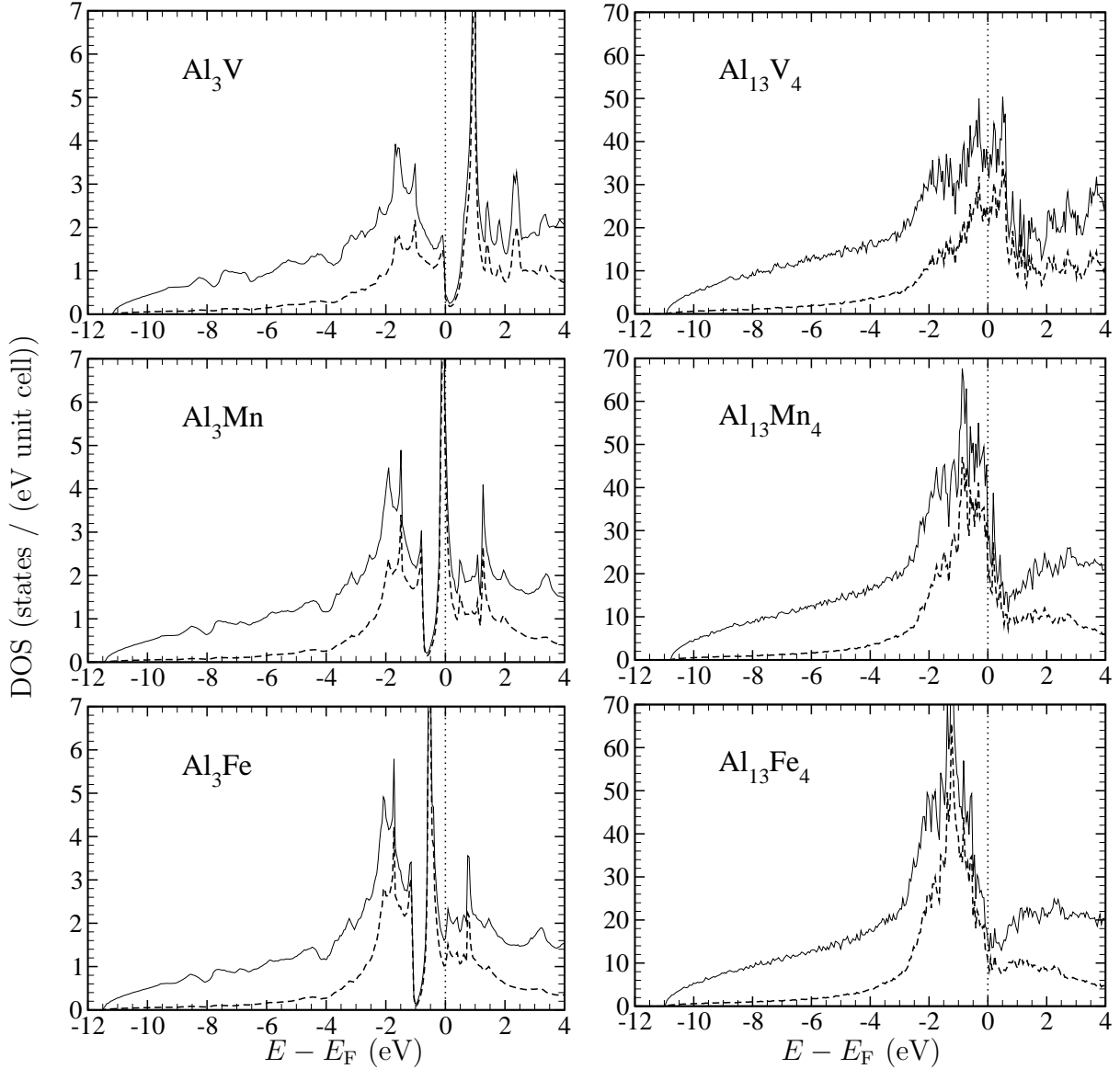

\begin{center}
\begin{tabular}{rcc}

\rotatebox[origin=c]{90}{DOS (states / (eV unit cell))}

&

\begin{minipage}[c]{7.3cm}
\begin{flushright}

\includegraphics[width=7.3cm]{Al3V_all_V.eps}~

\vspace{0.1cm}
\includegraphics[width=7.3cm]{Al3Mn_all_Mn.eps}~

\vspace{0.1cm}
\includegraphics[width=7.3cm]{Al3Fe_all_Fe.eps}~

\end{flushright}
\end{minipage}
&
\begin{minipage}[c]{7.5cm}

\includegraphics[width=7.4cm]{Al13V4_dos8_all_V.eps}

\vspace{0.1cm}
\includegraphics[width=7.4cm]{Al13Mn4_dos2_all_Mn.eps}

\vspace{0.1cm}
\includegraphics[width=7.4cm]{Al13Fe4_dos5_all_Fe.eps}

\end{minipage}
\\

 & ~~~~~$E-E_{\rm F}$ (eV) & $E-E_{\rm F}$ (eV)\\

\end{tabular}

\caption{LMTO (line) total DOSs
and (dashed line) TM DOSs of tetragonal
I4$/$mmm $\rm Al_3$$TM$
(structure and atomic positions of
$\rm Al_3V$);
and
monoclinic C2$/$m $\rm Al_{13}$$TM_{\rm 4}$
(structure and atomic positions
of $\rm Al_{13}Fe_{4}$)
for $TM={\rm V}$, Fe, Mn. \cite{PMS}
\label{LMTO_Comp_Al3V_Al13Fe4}}
\end{center}
\end{figure}

Considering first the case $TM={\rm Ru}$,
from an ab initio calculated equilibrium volume,
the heat of formation has been determined,
and it is shown that
monoclinic $\rm Al_{13}Ru_{4}$ is indeed more stable than
$\rm Al_{3}Ru$~\cite{Duc95},
as reported
in Al-Ru binary phase diagram~\cite{Massalski90}.
Moreover,
a detailed investigation of the link
between the density of
states and the stability using ab initio calculations
and a ``frozen-potential''
approach,
shows the
importance of the position of \ef \cite{Duc95}.
Indeed,  \ef is located near
a  peak in the DOS of $\rm Al_{3}Ru$, whereas it is located
in a the pseudogap in the DOS of $\rm Al_{13}Ru_{4}$~\cite{Duc95}.
In the former structure,
the central peak which is predominantly the
{\it ``non-bonding''} Ru $4d$, disappears from
the DOS of $\rm Al_{13}Ru_{4}$ where there are now
only bonding and anti-bonding hybrid $sp$(Al) and $d$(Ru) states between
the nearest neighbors.

We have  analyzed the non-bonding peak in Al-$TM$ alloys
across the transition-metal $3d$ series using a model taking into
account the diffraction by the Bragg plane in $spd$
compounds \cite{EurophysLet93,GuyPRB95}.
In the figure \ref{LMTO_Comp_Al3V_Al13Fe4},
the LMTO DOS of tetragonal
$\rm Al_{3}$$TM$
and monoclinic
$\rm Al_{13}$$TM$$_{\rm 4}$ are compared for
$TM = {\rm V}$, Mn and Fe.
With Mn and Fe the results are  similar to those
with Ru~\cite{Duc95},
whereas the situation is the opposite with V.
Indeed a non-bonding peak is observed in the DOS of
monoclinic $\rm Al_{13}V_4$
whereas \ef is near the pseudogap in
the DOS of tetragonal
$\rm Al_3V$.

\begin{figure}[t]
\begin{center}
\includegraphics[height=11cm]{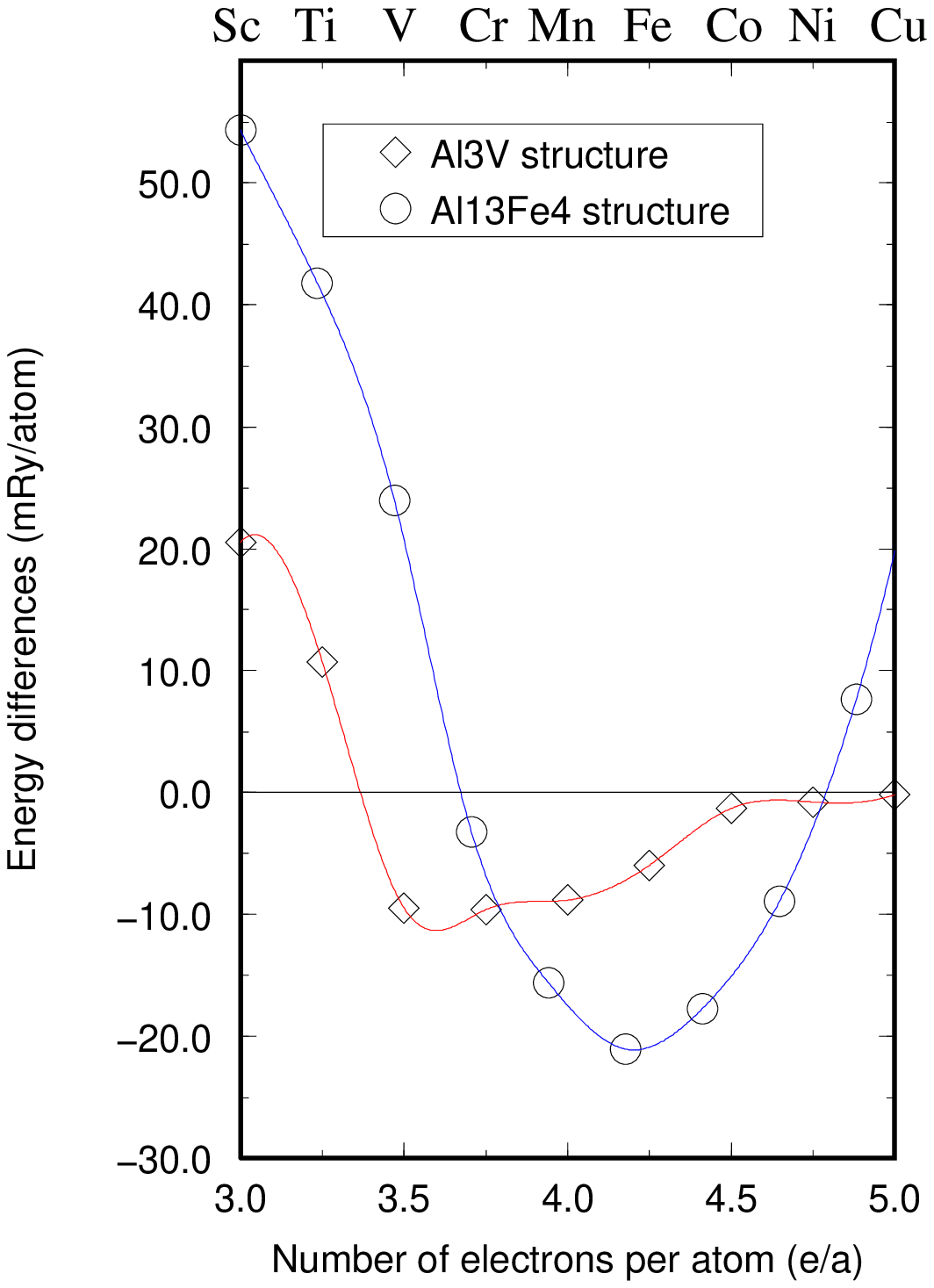}
\caption{Structural energy difference between the
tetragonal $\rm Al_{3}$$TM$
(I4/mmm $\rm Al_3V$ structure) and the
monoclinic $\rm Al_{13}$$TM$$\rm _{4}$
(C2/m $\rm Al_{13}Fe_4$ structure).
\cite{Duc95}}
\label{Duc95_intermetallicFig4}
\end{center}
\end{figure}

By using a Rigid Band Approximation
within local Density Functional
theory,
the effect of the number of electrons per atom $e/a$
on the relative stability  has been
studied \cite{Duc95}.
This shows (figure~\ref{Duc95_intermetallicFig4})
that transition-metal trialuminides goes
from the tetragonal $\rm Al_3$$TM$ structure
to the monoclinic $\rm Al_{13}$$TM_{\rm 4}$ structure
as a function of the electron
per atom ratio.
The $\rm Al_3$$TM$ structure is  more stable for
transition-metal trialuminides with $TM$ at the
beginning of the $d$ series (Sc, Ti, V, Y, Zr, Nb,
La, Hf, Ta)
whereas $\rm Al_{13}Fe_{4}$ structure is more stable
for
transition-metal trialuminides with $TM = {\rm Mn}$, Fe, Co, Ni
Tc, Ru, Rh, Pd, Re, Os, Ir and Pt~\cite{Duc95}.

These theoretical predictions of the relative stability
of the transition-metal trialuminides
between the simple tetragonal $\rm Al_3$$TM$ structure
and the more complex monoclinic $\rm Al_{13}$$TM_{\rm 4}$ structure,
agrees with the Hume-Rothery condition for stabilization.
Concerning the DOS, it results that the consequences of the
Hume-Rothery rules are the same as for $sp$ electron compounds:
The most stable phases are those for which \ef
is located in a pseudogap of the total DOS.

\subsection{Effective TM--TM medium-range interactions
stabilize Al--TM alumindes}

Zou and
Carlsson \cite{ZouPRL93} remarked first that
in many Al--Mn crystals approximants and quasicrystals the Mn--Mn
pair distances correspond to local minima of
$\Phi_{\rm Mn\textrm-Mn}(r)$ up to $\rm 10\,\AA$ and more.
That suggests the importance of the effective interaction
over medium-range order
for the stabilization of complex structure in Al(rich)--Mn
phase diagram.

The
internal energy $U$
is calculated
of all TM atoms in
Al(Si) host.
The reference energy, $U=0$, is those of one TM in the Al matrix
(TM impurity).
It corresponds to the energy for
the system where TM atoms are in the Al(Si) host
and
all TM--TM distances are  equal to infinity.
So, $U$ is defined as the energy needed to built the
structure from isolated TM atoms in the Al(Si) host.
It appears as a {\it ``structural energy''} of TM sub-lattice.
For a crystal,
$U_{\rm TM}$ per TM atom is:
\begin{eqnarray}
U_{\rm TM} = \frac{1}{\mathcal{N}_{\rm TM}} 
\sum_{\rm TM(k)} \alpha_k U_{\rm TM(k)}\;,
\label{EquationEStruturale}
\end{eqnarray}
The sum is on non-equivalent TM(k) Wyckoff sites. $\alpha_k$ is
the number of TM(k) atoms in a unit cell, and 
$\mathcal{N}_{\rm TM}$, $\mathcal{N}_{\rm TM}=\sum_{k} \alpha_k$, 
the number of TM atoms in a unit cell. 
$U_{\rm TM(k)}$ is the part of the structural energy
due to each TM(k) atom. 
$U_{\rm TM(k)}$ is
computed from the TM--TM pair interaction:
\begin{eqnarray}
{U}_{TM(k)}=\frac{1}{2}\sum_{i\neq k}
\Phi_{\rm TM\textrm{-}TM}(r_{ki})~e^{-\frac{r_{ki}}{L}}\;.
\label{EquationSumE}
\end{eqnarray}
$r_{ki}$ is the distance
between an atom TM$_k$ (on TM(k) Wyckoff site)
and TM$_i$ atoms. 
The sum is over all TM$_i$ atoms (TM$_i \neq$ TM$_k$).
${U}_{TM(k)}$ takes the
different values for all TM atom located on the different
Wyckoff sites.
The effective TM--TM interaction $\Phi_{TM-TM}$ is calculated
from the DOS of two TM in the Al matrix \cite{PMS}.
$L$ is mean-free path due to static disorder or/and by
phonons \cite{PMS}.
It is difficult to estimate and depends on the structural quality
and temperature. For metallic crystals and for approximants of quasicrystals, 
it should be larger
than $\rm 10\,\AA$ \cite{PMS}.

\begin{figure}[t!]
\begin{center}
\includegraphics[width=10cm]{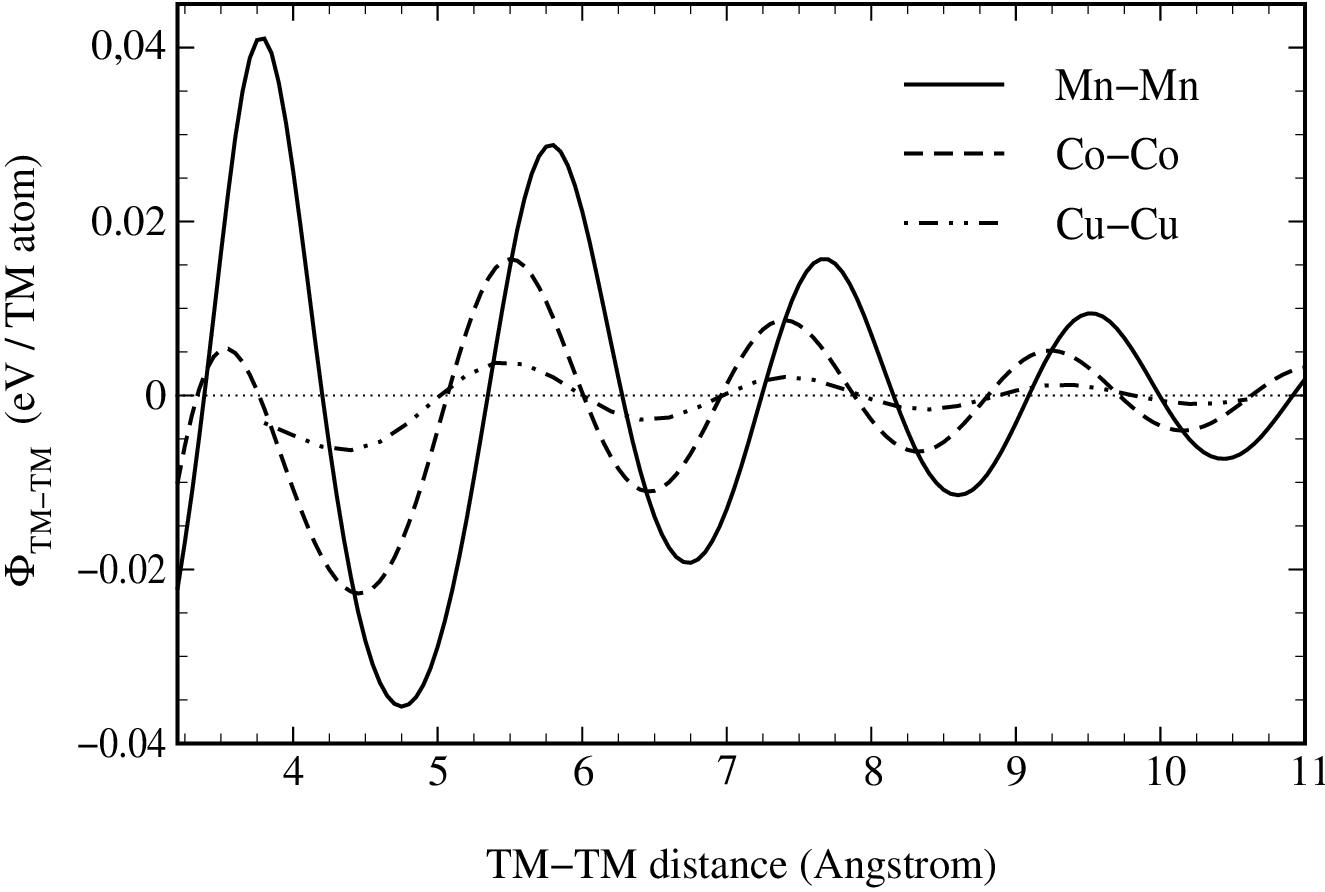}
\caption{
Effective medium-range pair interactions between two TM atoms
in a jellium (free states) simulating the Al matrix.
}
\label{FigPot_Mn_Co_Cu}
\end{center}

\begin{center}
\includegraphics[width=13cm]{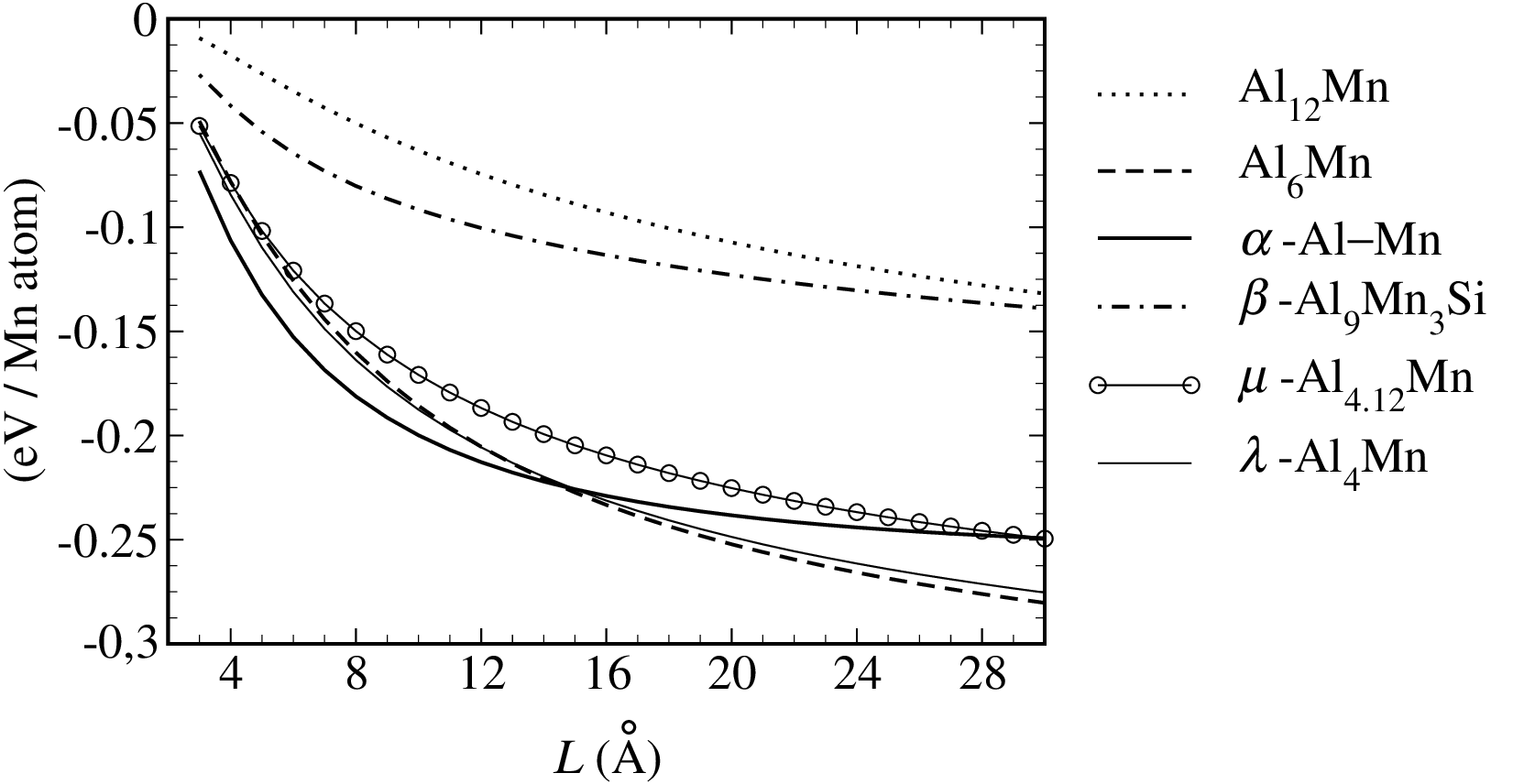}
\caption{Structural energy the Mn sub-lattice
of Al--Mn crystals.
The first-neighbors Mn--Mn contributions are not
include in the calculations. \cite{PMS}}
\label{E_pair_TMS}
\end{center}
\end{figure}

In  o-$\rm Al_6Mn$, $\rm Al_{12}Mn$ crystals
and
$\alpha$-Al--Mn--Si  approximant,
there is only one Mn Wyckoff position
therefore $U=U_{\rm Mn}$.
In these phases there is no Mn first-neighbors.
For phases that contain first-neighbor Mn pairs,
an energy $U'_{TM(i)}$ is calculated
without including the first-neighbor
TM--TM terms in the sum (\ref{EquationSumE}).
The corresponding $U'$ is the part of the structural energy
of TM sub-lattice that comes only from effective medium-range
TM--TM interactions.
For complex phases $\mu$-$\rm Al_4Mn$,
$\lambda$-$\rm Al_4Mn$
and $\rm Al_3Mn$,  $U'_{\rm Mn(i)}$ for different
Wyckoff Mn sites
have qualitatively the same
behavior;
thus, we present  their average
structural energy of the Mn sub-lattice
$U'=\langle U'_{\rm Mn(i)}\rangle_i$.
Figure \ref{E_pair_TMS} shows that the structural energy
of the Mn sub-lattice is always negative,
therefore the Mn--Mn
interactions over medium-range distances
contribute to stabilize these
phases.

In the liquid the loss of the medium-range order and the small
value of $L$ lead to $U$ close to zero. But it is still
negative~\cite{Guy02_EcoleChemnitz}, in agreement
with the fact that short-range and medium-range order do not
disappeared completely~\cite{SimonetPRB02,Schenk04}.

This approach is consistent which a Hume-Rothery
stabilization. Indeed, the Hume-Rothery mechanism
can be
expressed in terms of atomic interaction in the real space
\cite{Blandin67,Hafner87}.

\section{Origin of the vacancy in hexagonal
$\beta$-$\rm Al_9Mn_3Si$ and $\varphi$-$\rm Al_{10}Mn_3$}

Interesting examples of Al--TM crystals, are given by the
almost isomorphic
stable $\rm \beta$-$\rm Al_9Mn_3Si$,
meta-stable $\rm \varphi$-$\rm Al_{10}Mn_3$ and
stable $\rm Al_5Co_2$ phases.
In 1952, Robinson suggested that these compounds
with similar structure
could be  understood as Hume--Rothery phases with
similar $e/a$ ratios in spite of different atomic
concentrations~\cite{Robinson52}.
Indeed,  a
band energy minimization occurs when the Fermi sphere touches
a pseudo-Brillouin zone, spanned by Bragg vectors
${\bf K}_p$ corresponding to intense peaks in the experimental
diffraction pattern.

In an hexagonal unit cell
($\rm P6_3/mmc$),
$\rm \beta$-$\rm Al_9Mn_3Si$ ($\rm \varphi$-$\rm Al_{10}Mn_3$)
contains 18 (20) Al, 2 (0) Si, and 6 (6) equivalent Mn
(Mn(1) on Wyckoff site (6h));
and
$\rm Al_5Co_2$ contains  20 Al, and 8 Mn
(6 Co(1) on site (6h), and 2 Co(0) on site (2d)).
A particularity of the atomic structure of $\beta$ and
$\varphi$ phases is the presence of a large vacancy Va
on site (2d). But in $\rm Al_5Co_2$, this site is
occupied by Co(0). This explains the difference
of stoichiometry.
It is thus interesting to understand
why this vacancy is maintained in
$\beta$ and $\varphi$ crystals?
Because the first-neighbor distances around Va in
$\beta$, $\varphi$
are close to those around Mn(1) in $\beta$, $\varphi$ and to
those around Co
in $\rm Al_5Co_2$,
the presence of
Va cannot be explained from steric encumbering.
Indeed,
the environment of the vacancy in $\beta$, $\varphi$ forms a
tri-capped trigonal prism
(3 Al(1) and 6 Al(2)).
A similar environment, with similar inter-atomic distances,
is found in
$\rm \mu$-$\rm Al_{4.12}Mn$ (Ref.~\cite{Shoemaker89}) and
$\rm \lambda$-$\rm Al_{4}Mn$ (Ref.~\cite{Kreiner97}).
But in $\mu$ and $\lambda$, these sites are occupied
by  Mn.

\subsection{ab initio study}

\begin{figure}[t!]
\begin{center}
\begin{tabular}{ccc}

\rotatebox[origin=c]{90}{(states / (eV unit cell))}
&
\begin{minipage}[c]{7.2cm}

\begin{flushright}
\includegraphics[height=4.5cm]{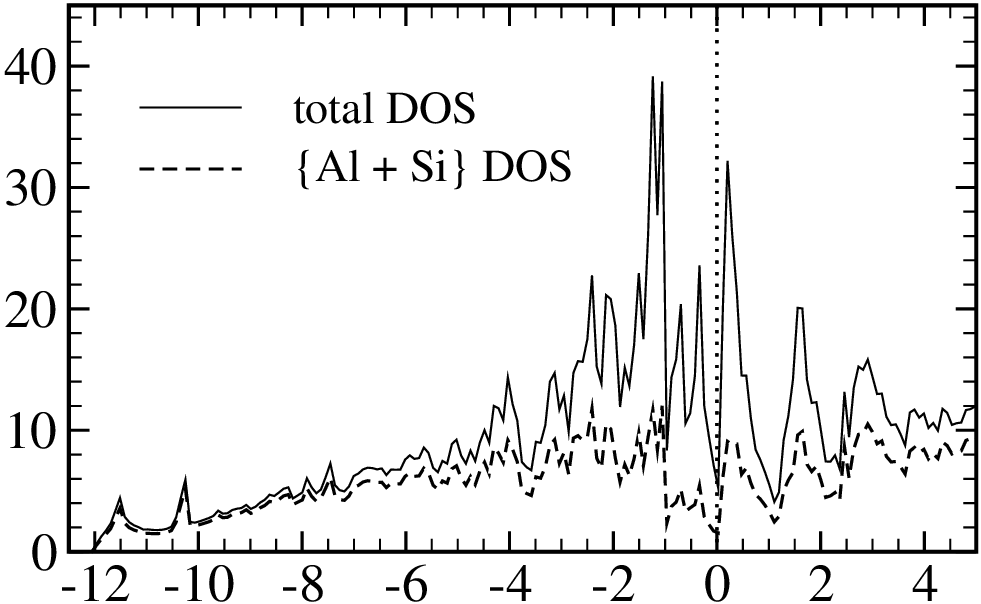}

\end{flushright}
\end{minipage}
\vspace{0.1cm}

&
\begin{minipage}[c]{7.2cm}

\begin{flushright}
\includegraphics[height=4.5cm]{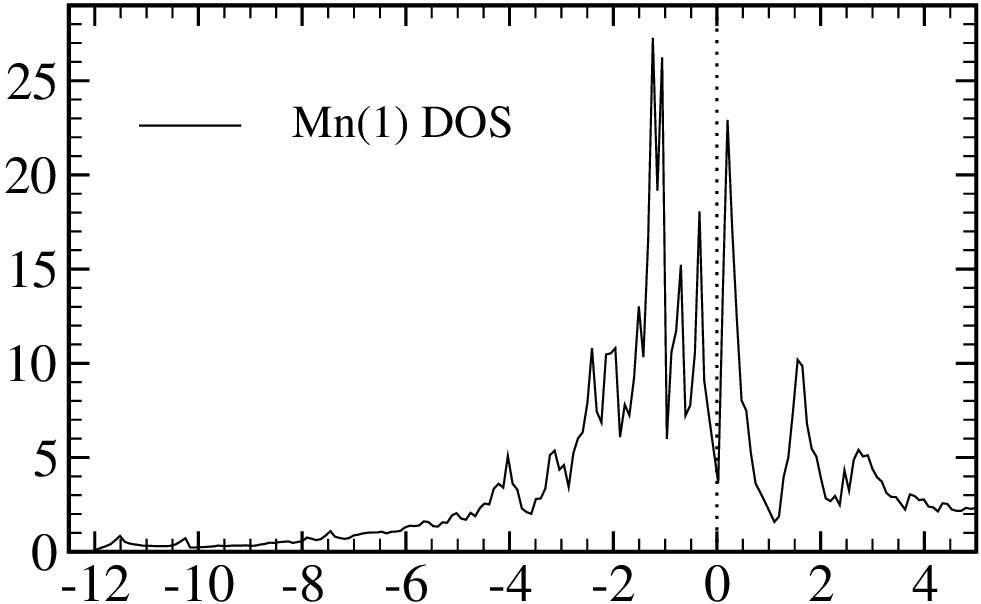}

\end{flushright}
\end{minipage}
\vspace{0.1cm}

\\

 & ~~$E-E_{\rm F}$ (eV) & ~~$E-E_{\rm F}$ (eV)\\
\end{tabular}
\caption{LMTO of hexagonal $\beta$-$\rm Al_9Mn_3Si$. \cite{Guy03}
\label{LMTO_beta}}
\end{center}
\end{figure}

\begin{figure}[t]
\begin{center}

\begin{tabular}{ccc}

 & ~~Hypothetical $\beta$-$\rm Al_9Mn_4Si$& ~~$\rm Al_5Co_2$\\

\rotatebox[origin=c]{90}{(states / (eV unit cell))}
&
\begin{minipage}[c]{7.2cm}

\begin{flushright}
\includegraphics[height=4.5cm]{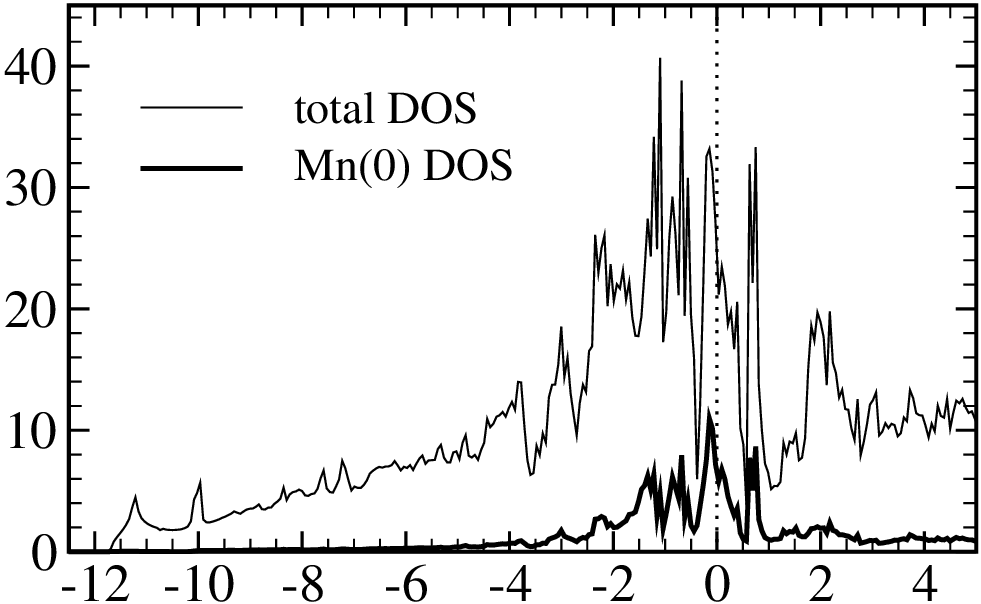}

\end{flushright}
\end{minipage}
\vspace{0.1cm}

&
\begin{minipage}[c]{7.2cm}

\begin{flushright}
\includegraphics[height=4.5cm]{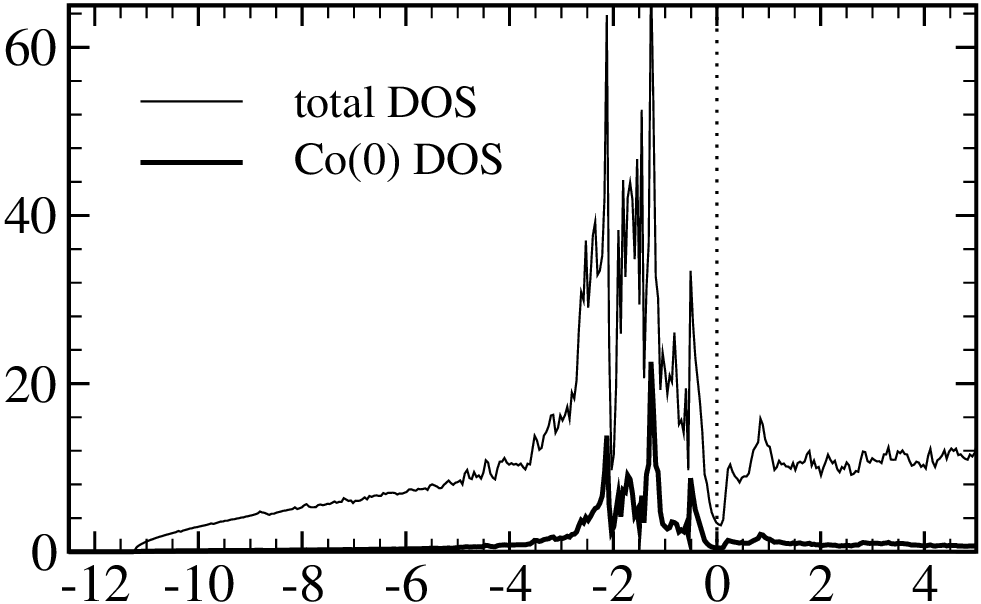}

\end{flushright}
\end{minipage}
\vspace{0.1cm}

\\

 & ~~$E-E_{\rm F}$ (eV) & ~~$E-E_{\rm F}$ (eV)\\
\end{tabular}
\caption{LMTO DOSs of the almost isomorphic hypothetical
$\beta$-$\rm Al_9Mn_3Si$, and actual $\rm Al_5Co2$. (see text)
\label{LMTO_hyp_beta}}

\vspace{0.5cm}
\begin{tabular}{ccc}

 & ~~Hypothetical $\beta$-$\rm Al_9Mn_4Si$& ~~$\rm Al_5Co_2$\\

\rotatebox[origin=c]{90}{(eV $/$ TM atoms)}
&
\begin{minipage}[c]{7.10cm}

\begin{flushright}
\includegraphics[height=4.42cm]{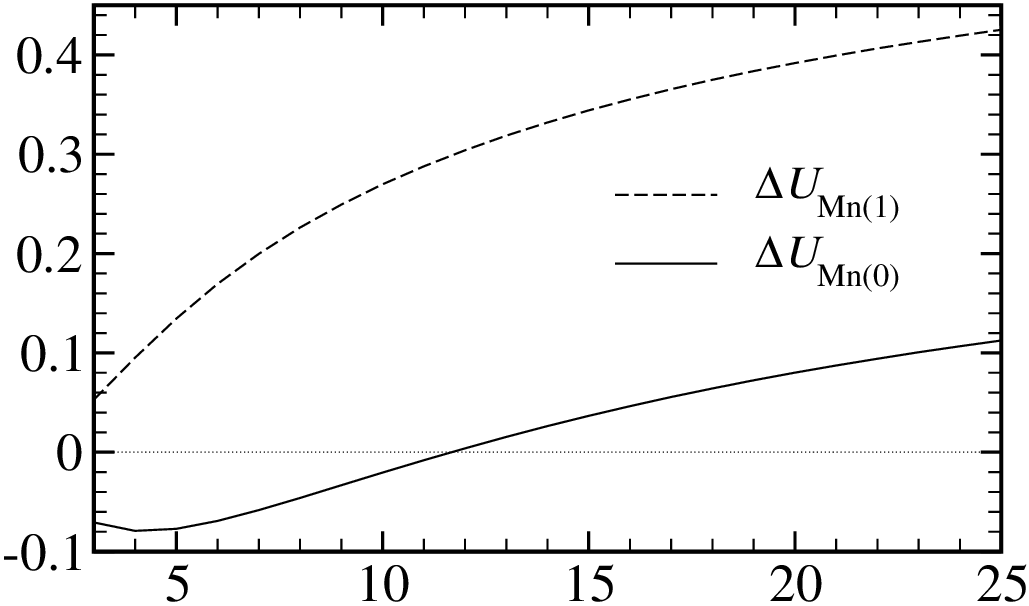}

\end{flushright}
\end{minipage}
\vspace{0.1cm}

&
\begin{minipage}[c]{7.2cm}

\begin{flushright}
\includegraphics[height=4.42cm]{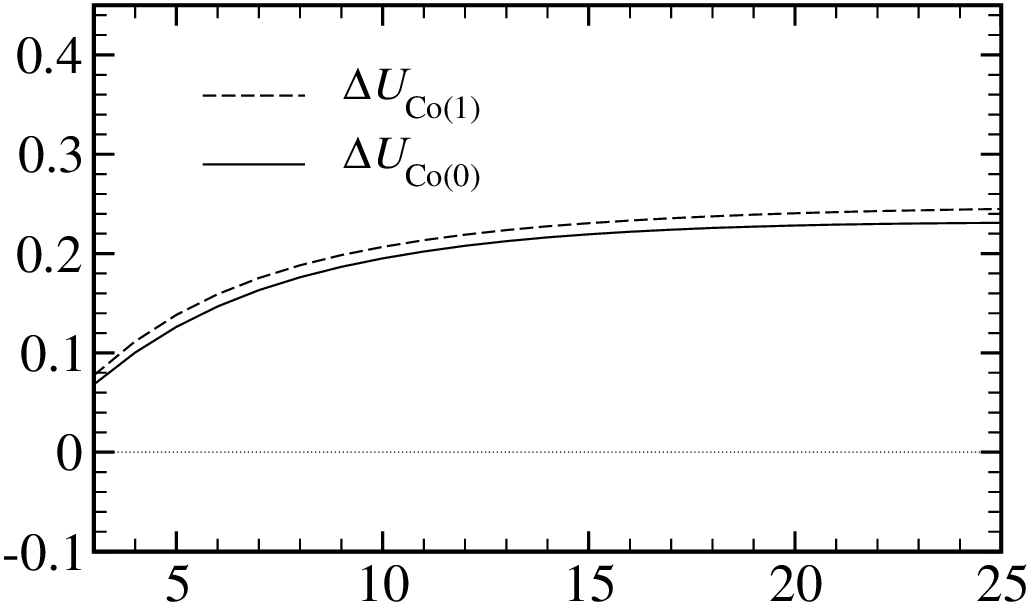}

\end{flushright}
\end{minipage}
\vspace{0.1cm}

\\

 & ~~$L$ (\AA) & ~~$L$ (\AA)\\
\end{tabular}
\caption{Variation of the structural energy $\Delta U$ due
to the effective TM--TM interaction in the almost isomorphic hypothetical
$\beta$-$\rm Al_9Mn_3Si$, and actual $\rm Al_5Co2$. (see text)
\cite{Guy03}
\label{DeltaE_hypo_beta}}

\end{center}
\end{figure}

The LMTO total DOS and the local \{Al $+$ Si\} DOS of
$\beta$-$\rm Al_9Mn_3Si$ are
shown in figure \ref{LMTO_beta}.
A pseudogap in the \{Al + Si\} DOS is clearly seen.
This large pseudogap  is
mainly characteristic of a
$p$ band at this energy,
but  the pseudogap in the total DOS is narrower.
In Ref. \cite{Guy03}, we have shown that the pseudogap
in $sp$ DOS is due to the scattering of $sp$ states by the Mn sub-lattice
(called Mn(1) sub-lattice) via a strong $sp$-$d$ hybridization.
To analyze the origin of the vacancy in $\beta$ phases,
we have performed calculation including a new Mn atom, called
Mn(0), on site (2d) in $\beta$-$\rm Al_9Mn_3Si$ structure.
Atomic positions and lattice parameters are those of
$\beta$-$\rm Al_9Mn_3Si$.
This hypothetical phase is named $\beta$-$\rm Al_{9}Mn_4Si$.
The absence of pseudogap in the total DOS
(figure \ref{LMTO_hyp_beta}) shows
a the great effect on Mn(0) which
is very different from the one of Mn(1).
Indeed Mn(1) (on site (6h))
creates the pseudogap in $\beta$-$\rm Al_9Mn_3Si$ DOS,
whereas
Mn(0) destroys it in
hypothetical  $\beta$-$\rm Al_9Mn_4Si$ total DOS.
Thus
Mn(0) on site (2d)  ``fills up'' the
pseudogap via the $sp$-$d$ hybridization;
whereas  Mn(1) on site (6h) enhances the pseudogap.
A similar result is obtained for hypothetical
$\varphi$-$\rm Al_{10}Mn_4$, by putting Mn(0) in place of
the vacancy in $\varphi$-$\rm Al_{10}Mn_3$ phases.
This illustrates clearly the non-trivial effect of the Mn positions
on the electronic structure
of $spd$ Hume-Rothery
alloys.

On the other hand, Co(0) on site (2d) increases the pseudogap
(figure \ref{LMTO_hyp_beta}) in $\rm Al_5Co_2$.

\subsection{Medium-range Mn--Mn interaction can induce vacancies in
the atomic structure}

To understand the origin of the vacancy (Va) in $\beta$ and $\varphi$
structure, it is not enough to analyze the local environment (first neighbor
environment). Indeed, the local environment around  vacancy
(table \ref{Tab_DistancesBeta}),
is very similar \cite{Guy03} to that of Co(O) in $\rm Al_5Co_2$ and
that of Mn(0) in
hexagonal
$\mu$-$\rm Al_{4.12}Mn$ \cite{Shoemaker89} and
$\lambda$-$\rm Al_4Mn$ \cite{Kreiner97}.
Therefore we have considered
the medium-range TM--TM interactions.

For phases containing several TM Wyckoff sites,
the effective TM--TM  interaction allows
to compare the relative stability of TM atoms on
different Wyckoff sites.
%
Considering the hypothetical $\beta$-$\rm Al_9Mn_4Si$
phase (see previous paragraph),
the variation, $\Delta {U}_{\rm Mn}$, of ${U}$
is determined
when
one Mn atom ($\rm Mn_i$) is removed from the structure and is
put as an impurity in an Al matrix:
\begin{eqnarray}
\Delta {U}_{\rm Mn_i}=-\sum_{k\,(k\neq i)}
\Phi_{\rm Mn\textrm{-}Mn}(r_{ik})~\textrm{e}^{-\frac{r_{ik}}{L}}\;.
\label{EquationSumDE}
\end{eqnarray}
$\Phi_{Mn\textrm{-}Mn}$ is the effective TM--TM interaction
(figure \ref{FigPot_Mn_Co_Cu});
$r_{ik}$, the $\rm Mn_i$--$\rm Mn_k$ distance; and $L$,
the mean-free path due to static disorder or/and
phonons.
Mn atoms on the same Wyckoff sites have the
same $\Delta {U}_{\rm Mn}$ value, Mn atoms on different Wyckoff sites
have different $\Delta U_{\rm Mn}$ values that can
be compared.
The most stable Mn Wyckoff sites correspond to the highest
$\Delta {U}_{\rm Mn(k)}$ values.
As previously, the
energy is calculated from equation
(\ref{EquationSumDE}) without the first-neighbor
Mn--Mn contributions in order to analyze
effects at medium-range order.

For hypothetical $\beta$-$\rm Al_9Mn_4Si$ (where Mn(0) replaces the
vacancy), one finds:
$\Delta {U}_{\rm Mn(1)}' > \Delta {U}_{\rm Mn(0)}'$
(figure \ref{DeltaE_hypo_beta}).
Mn(0) in (2d)
is therefore less stable than Mn(1) in (6h) in the hypothetical
$\beta$-$\rm Al_9Mn_4Si$,
in agreement with the fact that
(2d) site is empty (vacancy) in $\beta$-$\rm Al_9Mn_3Si$ phase.
A similar result was obtained for
the hypothetical
$\varphi$-$\rm Al_{10}Mn_4$.

$\rm Al_5Co_2$ phase is almost isomorphic of
$\beta$ and $\varphi$ phases, but
there is a Co site (Co(0)) corresponding to the vacancy
of $\beta$ and $\varphi$.
$\Delta {U}_{\rm Co(0)}$, calculated
with the effective Co--Co interaction, is almost equal
to $\Delta {U}_{\rm Co(1)}$ (figure \ref{DeltaE_hypo_beta}),
thus
Co(0) in (2d) is as stable as Co(1) in (6h).
This justifies why
no vacancy  exists in $\rm Al_5Co_2$.

\begin{table}
\begin{center}
\caption{\label{Tab_DistancesBeta} Inter-atomic distances
around the site (2d)
in $\beta$-$\rm Al_9Mn_3Si$,
$\varphi$-$\rm Al_{10}Mn_3$ and  $\rm Al_5Co_2$.
TM(1) is either Mn(1) or Co(1).
$X$ corresponds to the vacancy Va in
$\beta$,  $\varphi$ phases,
and to Co(0) in $\rm Al_5Co_2$.}
\begin{tabular}{|l|c|c|c|}
\hline
 Neighbors & \multicolumn{3}{c|}{Distances ($\rm \AA$)}  \\
\cline{2-4}
& $\rm \beta\,Al_9Mn_3Si$ & $\rm \varphi\,Al_{10}Mn_3$
& $\rm Al_5Co_2$  \\
\hline
 3 Al & 2.72 & 2.77 & 2.61 \\
 6 Al & 2.23 & 2.29 & 2.35 \\
 6 TM(1)$\rm ^{*}$& 3.81 & 3.82 & 3.86 \\
\hline
\end{tabular}\\
~~$^{*}$ Atom in (2d) and TM(1) are not first-neighbor. \\
\end{center}
\end{table}

The difference between $\varphi$, $\beta$ phases, and $\rm Al_5Co_2$
phase, is understood by considering the TM--TM effective
interaction $\Phi_{\rm TM\textrm{--}TM}$
(figure \ref{FigPot_Mn_Co_Cu}).
In
$\varphi$, $\beta$ phases,
the environment of Va  contains two Mn atoms at the distance
{3.8\,\AA} (table \ref{Tab_DistancesBeta}).
Similarly, in $\rm Al_5Co_2$ the environment of Co(0)
contains two Co at the distance {3.8\,\AA}.
$\Phi_{\rm Mn\textrm{--}Mn}(3.8\,{\rm \AA})>0$, then
{3.8\,\AA} corresponds to an unfavorable Mn--Mn distance;
whereas
$\Phi_{\rm Co\textrm{--}Co}(3.86\,{\rm \AA})<0$, then
{3.86\,\AA} corresponds to a favorable  Co--Co distance.

Finally, we have compared \cite{Guy03} the case of $\beta$, $\varphi$ phases with
the hexagonal
$\mu$-$\rm Al_{4.12}Mn$ \cite{Shoemaker89} and
$\lambda$-$\rm Al_4Mn$ \cite{Kreiner97} that contain
an atomic site with a very similar environment to that of Va in $\beta$, $\varphi$.
From X-ray data, this site is occupied by a Mn (Mn(0)) in $\mu$, $\lambda$.
That difference with $\beta$ $\varphi$ cannot be explained in term of
local environment whereas the medium-range Mn--Mn interaction explains
it \cite{Guy03}.
Indeed, in $\mu$, $\lambda$ there is no Mn(0)--Mn distance
equal to 3.8\,\AA, but the first Mn(0)--Mn distance is 4.8\,$\rm \AA$ which
corresponds to a favorable Mn--Mn spacing
(figure \ref{FigPot_Mn_Co_Cu}).

\section{Magnetism of Al(Si)--Mn phases}

\subsection{Introduction and ab initio studies of magnetism}

The Mn impurity
in Al host is close to a magnetic~$/$ non-magnetic transition.
But the situation in Al--Mn alloys is rather different as
most of Mn atoms are non-magnetic.
Indeed, it is found experimentally that in simple crystals
Mn is non-magnetic, and in complex phases and approximants
only a small proportion of Mn is magnetic
(F. Hippert, V. Simonet et al.
\cite{Hippert99_JPCM,Hippert00_MRS,Hippert_Liquide,Simonet98_PRB,Hippert03},
C. Berger and J.J. Pr\'ejean et al.~\cite{Berger94,Prejean02})

\begin{figure}[t]
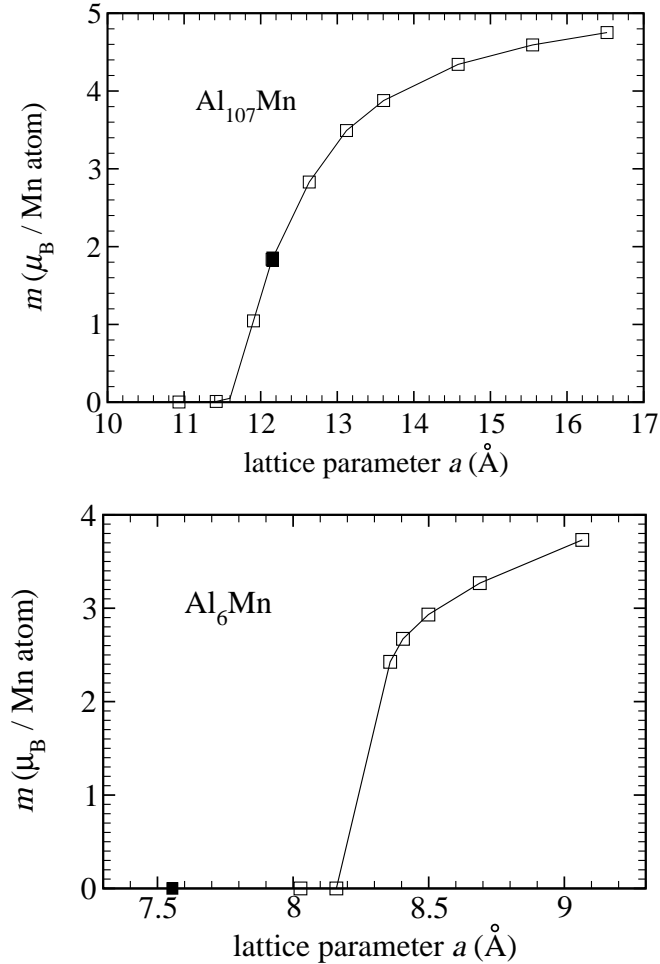

\begin{center}
\includegraphics[width=8.5cm]{Dilatation_Al107Mn_041115_2.eps}

\vspace{0.25cm}
\includegraphics[width=8.5cm]{Dilatation_Al6Mn.eps}~~

\caption{Local moment on Mn atoms in a cubic $\rm Al_{107}Mn$ model
(impurity model) and o-$\rm Al_6Mn$,
versus the lattice parameter of the cubic lattice
(isotropic dilatation).
The black square corresponds to the experimental
lattice parameter of f.c.c. Al and o-$\rm Al_6Mn$, respectively.
Lines are guides for the eyes.
\label{LMTO_dilatation}
}
\end{center}
\end{figure}

The LMTO calculations confirm that $\rm Al_{12}Mn$, o-$\rm Al_6Mn$,
$\varphi$-$\rm Al_{10}Mn_3$, $\beta$-$\rm Al_9Mn_3Si$,
$\alpha$-Al--Mn--Si, $\rm Al_{11}Mn_4$,
$\rm Al_{13}Mn_4$ ($\rm Al_{13}Fe_4$ structure)
are non-magnetic.
The fact that  Mn atoms are not close to
a magnetic~$/$ non-magnetic transition is shown by
the effect of the  dilatation
(figure~\ref{LMTO_dilatation}).
For instance in the case of $\rm Al_6Mn$
an isotropic dilation of lattice parameters of
$\sim 8$\% is necessary
to reach the magnetic transition,
whereas an Mn in substitution in the Al matrix (Al f.c.c.)
is on the magnetic~$/$ non-magnetic transition.

In the magnetic complex Al--Mn--(Pd)--(Si) phases,
LMTO calculations
show that some Mn are non-magnetic
and other Mn  are magnetic
(Al--Mn approximants \cite{Hafner98a},
Al--Pd--Mn approximants \cite{Krajci98},
1/1-$\rm Al_{65.9}Pd_{12.2}Mn_{14.6}Si_{7.3}$ \cite{Hippert99_JPCM},
$\mu$-$\rm Al_{4.12}Mn$ \cite{Duc03}).
These calculations
suggest also that the
magnetic Mn atoms are located on Mn sites less stable than
the Mn sites occupied by  non-magnetic Mn.

Up to now, most of the theoretical studies have focused on the role
of the local environment of the Mn atoms to explain the occurrence of
localized magnetic moment
like in the case of Mn impurity in the Al matrix
(see Refs. in \cite{PRL00}).
In particular, clusters calculations have
shown that the local symmetry and the
first-neighbor
distance
can have a strong influence. Vacancies, Mn
pairs or Mn triplets or Mn quadriplets are also often invoked to explain
magnetic moments
\cite{Hippert99_JPCM,Simonet98_PRB,Hippert03}.
However, in crystals and
quasicrystals, most of Mn atoms are non-magnetic in spite of various
environments including  pairs, triplets and quadriplets.
It is therefore expected that local environment properties are
not enough to understand the existence of localized magnetic moment
in Al--Mn compounds \cite{PRL00}.

Theoretical works have focused on the occurrence
of local moments in a series of Al--Mn
alloys~\cite{Bratkovsky95,Hafner98a}.
The authors conclude that the
crystal o-$\rm Al_6Mn$ is non-magnetic because of a pseudogap in the
local density of states (DOS) at the Fermi energy which is of a
Hume--Rothery type. Because of the chemical disorder, a solid solution
at the same concentration (on an Al f.c.c. lattice without
relaxation) presents a very different electronic DOS, without a
pseudogap at the Fermi energy \cite{Hafner98a}. In this solution the
Mn atoms are magnetic.
In the Hume--Rothery mechanism, the
pseudogap contributes to the stabilization of crystals,
thus
magnetic state of Mn atoms is related to the stabilization mechanism.

An illustration of the effect of a strong effective Mn--Mn interaction
on magnetism has been given by an LMTO calculation
on  $\beta$-$\rm Al_9Mn_3Si$ and $\varphi$-$\rm Al_{10}Mn_3$
\cite{PRL00}.
The hexagonal unit cell of these phases, contains two isolated
Mn-triplets.
The Mn atoms belonging to the same triplet are
first-neighbors, but each Mn-triplet is
surrounded by Al(Si) atoms only.
LMTO calculations show that
the Mn-triplets are non-magnetic.
In order
to determine the role of the effective Mn--Mn interaction on this
result, we performed a calculation
for an hypothetical
$\beta$-$\rm Al_9Mn_{1.5}Cu_{1.5}Si$ phase, constructed from
$\beta$-$\rm Al_9Mn_3Si$ by replacing one Mn-triplet by a Cu-triplet in
each cell.
The Cu has no medium-range interaction as its $d$ orbitals are
full and it has almost the same number of valence ($sp$) electrons as
Mn.
Thus the Fermi energy is essentially unchanged as well as the
local environment of the Mn-triplet.
Yet the LMTO results show a
magnetic moment equal to $\rm \sim1\,\mu_B$ on each Mn in
$\beta$-$\rm Al_9Mn_{1.5}Cu_{1.5}Si$
(the 3 Mn atoms in a triplet are almost
equivalent with a ferromagnetic spin orientation).
This is a proof
that the magnetic state of an Mn atom is sensitive to Mn atoms at a
distance of $\rm \sim 5\,\AA$ (distance between two
Mn-triplet in $\beta$-phase).
The energy of formation of magnetic
moments in $\beta$-$\rm Al_9Mn_{1.5}Cu_{1.5}Si$ is $-0.046\,\rm eV$
per triplet.
Similar results are obtained
for $\varphi$-$\rm Al_{10}Mn_3$.
These results confirm that an magnetic state is expected for an isolated
Mn-triplet in Al matrix, but in $\beta$ and $\varphi$,
a strong
inter-triplets Mn--Mn interaction
maintains a non-magnetic state.

\subsection{The magnetic Mn--Mn effective interaction}

The magnetic effective Mn--Mn interaction in Al(rich) alloys
is calculated as
follows \cite{PRL00}.
The $d$ orbitals of Mn atoms are coupled to free
states (Al  states mainly) but we neglect the direct $d$--$d$
coupling. The local magnetic moment is treated in a mean-field type
approach as in band-structure calculations. That is one neglects
spin fluctuations effect.
We now consider the energy $E$ of 2 Mn
atoms ($\rm Mn_1$ and $\rm Mn_2$) in an Al matrix which is simulated
by the jellium (free electron). $E$ is a function of the
$\rm Mn_1$--$\rm Mn_2$ distance,
$r_{12}$, and of the moments $m_1$ and $m_2$ carried
by the 2 Mn, respectively. By convention we choose
$E(r_{12}=\infty,m_1=m_2=0)=0$.
Then one may write \cite{PRL00}:
\begin{eqnarray}
E(r_{12},\overrightarrow{m_1},\overrightarrow{m_2}) = E_1(m_1) +
E_1(m_2)+
\Phi_{\rm Mn\textrm{--}Mn}
(r_{12},\overrightarrow{m_1},\overrightarrow{m_2})
\label{EnergieTotale}
\end{eqnarray}
$E_1(m_i)$ is the energy of one $\rm Mn_i$ impurity in an Al matrix
(Virtual Bound State), it does not depend on the position of
$\rm Mn_i$.
$\rm \Phi_{Mn\textrm{--}Mn}$ is an effective Mn--Mn
interaction which is mediated by
the $sp$ states,
and it depends on Mn--Mn distance and the moments carried by the Mn atoms.

As Mn impurity is close to the magnetic transition, $E_1(m)$
is small ($|E_1(m)|\lesssim 0.05$\,eV). The value of $E_1(m)$,
has been  estimated from LMTO
calculation  for the concentration $\rm
Al_{107}Mn$ (Mn atoms substituting Al atoms in a Al f.c.c. lattice,
without inter-atomic distance relaxation):
Mn is found magnetic with
a $\rm 1.8\,\mu_B$ moment and the gain in energy for one Mn atom to
become magnetic is $E_1(m=1.8\,\mu_B)\simeq-0.05$\,eV.

\begin{figure}[t]
\begin{center}
\includegraphics[width=10cm]{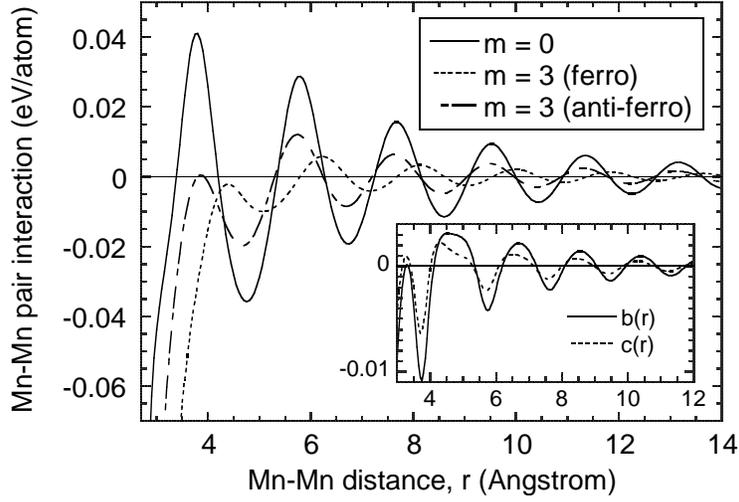}
\caption{Effective Mn--Mn interaction
$\rm \Phi_{\rm Mn\textrm{--}Mn}$ in Al
host (free electrons).
Non-magnetic case ($\rm m_1=m_2=0$); magnetic case: the two
moments ($\rm m_1=m_2=3\,\mu_B$) alignment is anti-ferromagnetic or
ferromagnetic. [Inset: Coefficients b(r) and c(r)
(equation(\ref{EqDeveloppement})), a(r)$=\rm \Phi(m_1=m_2=0)$.]}
\label{LabFigure1}
\end{center}
\end{figure}

We have calculated $\Phi$ for two magnetic
Mn atoms (2
$d$ orbitals) in free electron matrix
(figure \ref{LabFigure1}) \cite{PRL00}.
A
remarkable feature is that most pronounced minima in the Mn pair interaction
correspond to non-magnetic Mn pairs for specific Mn--Mn spacings.
Therefore the most stable Mn
sites should be occupied by non-magnetic Mn.
When Mn are  in
a less stable position, it could be magnetic. This conclusion is
consistent with experimental results from which most of the Mn atoms
are non-magnetic, whereas only particular sites (or defect)
could lead to
magnetic Mn.

For small moments, the Mn pair interaction energy can be developed
as:
\begin{eqnarray}
\Phi_{\rm{Mn\textrm{--}Mn}}(r_{12},m_1,m_2)
= a(r_{12}) + \frac{1}{2}\;b(r_{12})\;(m_1^2 + m_2^2)
+c(r_{12})\;\overrightarrow{m_1}\cdot\overrightarrow{m_2} + {}\cdots
\label{EqDeveloppement}
\end{eqnarray}
The $a(r)$ terms is the Mn--Mn interaction for non-magnetic Mn atoms
($m_1=m_2=0$) (figure \ref{FigPot_Mn_Co_Cu}).
The $c(r)$ term is the RKKY exchange interaction between the two
spins.
The $b(r)$ term plays a central
role in our study of the existence of local magnetic moments
in a non-magnetic environment.
As an example, let us consider the magnetic moment
$m_1$ of an atom $\rm Mn_1$ interacting with a non magnetic
$\rm Mn_2$ ($m_2=0$), in the Al matrix.
We neglect $E_1(m_1)$ which is small.
The interaction
between $\rm Mn_1$ and $\rm Mn_2$ depends on the magnetic
moment $m_1$  carried by $\rm Mn_1$ via the term
$b(r_{12})$:
$\Phi_{\rm{Mn\textrm{--}Mn}}(r_{12},m_1=0,m_2)
= a(r_{12}) + \frac{1}{2}\;b(r_{12})\;m_1^2$.
The minimum of $\Phi_{\rm{Mn\textrm{--}Mn}}$ (stable) is reached for
a $m_1$ value that depends on the sign of $b(r_{ij})$:
For $b(r_{ij})<0$, stable $\rm Mn_1$ is magnetic, whereas for
$b(r_{ij})>0$ stable $\rm Mn_1$ is non-magnetic.
Therefore, the term
$b(r)$ implies that the formation of a magnetic moment is sensitive to the
presence of other non-magnetic Mn.


We now consider the general case of more than 2 Mn atoms in Al matrix.
Starting from non-magnetic case, we calculate the formation energy,
$\Delta \mathcal{E}_i$, for the moment, $m_i$, on the atom $\rm Mn_i$:
\begin{eqnarray}
\Delta \mathcal{E}_i = E_{1}(m_i) +
\mathcal{B}_{i}m_i^2 ~~{\rm with}~~
\mathcal{B}_i=\sum_{j} \frac{b(r_{ij})}{2} e^{-\frac{r_{ij}}{L_0}}.
\label{eqB_i}
\end{eqnarray}
$r_{ij}$ is the $\rm Mn_i$--$\rm Mn_j$ distances.
$E_1$ include all first neighbors effects.
The sum $\rm {\cal B}_i$ takes into account the medium-range
effects of the Mn--Mn interaction.  When $|{\cal
B}_im_i^2|>|E_1(m_i)|$,
the sign of
${\cal B}_i$ determines the magnetic states of $\rm Mn_i$:
\begin{eqnarray}
{\cal B}_i > 0 & \Longrightarrow &
{\rm Mn_i~is~non\textrm{-}magnetic}~(m_i=0)\\
{\cal B}_i < 0 & \Longrightarrow & {\rm Mn_i~is~magnetic}~(m_i\neq 0)
\end{eqnarray}

\begin{figure}[t]
\begin{center}
\includegraphics[width=10cm]{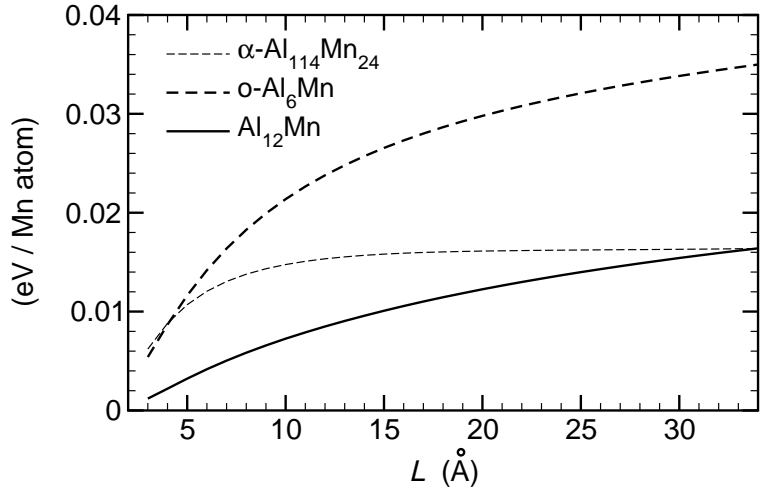}
\caption{
$\mathcal{B}_{{\rm Mn}}(L)$ for
$\rm \alpha$-$\rm Al_{110}Mn_{24}$,
o-$\rm Al_6Mn$,
and $\rm Al_{12}Mn$ \cite{PMS}.
In these crystals, Mn atoms are not first-neighbor.}
\label{FigChapMagBi_crystaux}
\end{center}
\end{figure}

For all Mn atoms in o-$\rm Al_6Mn$, $\rm
\alpha$-$\rm Al_{73}Mn_{17}Si_{10}$, $\rm \beta$-$\rm Al_9Mn_3Si$,
$\rm Al_3Mn$
phases, we found $\rm {\cal B}_i>0.015$eV when $L_0>20$\,$\rm \AA$
(figure \ref{FigChapMagBi_crystaux}).
Assuming $|E_1|\leq0.05$\,eV for $m\simeq2$\,$\rm \mu_B$,
$\Delta \mathcal{E}$
is minimized when Mn are non-magnetic, as found experimentally.

In liquid phases the situation is completely different. Because of
the loss of the medium range order, $\rm {\cal B}_i\simeq0$. Then
thermal expansion, displacements of the atoms and spin fluctuations,
should favor a non zero average moment as found
experimentally \cite{Hippert_Liquide} and by numerical
simulations \cite{Bratkovsky95,Hafner98a}.

We have also shown that $\rm {\cal B}$ interaction between atoms of
different Mn triplets plays a central role
in $\beta$-$\rm Al_9Mn_3Si$ \cite{PRL00,PMS}.
This interaction forbids the occurrence
of Mn moments in triplets, whereas a single Mn triplet in an Al matrix
should be magnetic.

\section{Electronic localization}

\subsection{Electronic transport}

T. Fujiwara {\it et al.} have first shown  that the electronic
structure of Al--TM approximants and related phases is
characterized by two energy scales
\cite{Fujiwara89,Fujiwara91,Fujiwara93,PRB94_AlCuFe,PRB94_AlCuCo}.
The largest, about $0.5 - 1$\,eV, is the width of the pseudogap
near the Fermi energy $E_{\rm F}$. It
is related to the Hume--Rothery stabilization via the scattering of
electrons by the TM sub-lattice because of a strong \spd
hybridization.
The smallest, less
than 0.1\,eV, is characteristic of the small dispersion of the
band energy $E({\bf k})$~\cite{Fujiwara93}.
This energy scale seems more
specific to phases related to the quasiperiodicity. The first
consequence as far as the transport is concerned is a small
velocity at Fermi energy
\begin{equation}
v(E_{\rm F})=\Bigg(\frac{\partial E}{\partial k}\Bigg)_{E = E_{\rm F}}.
\end{equation}
From LMTO calculations
the Bloch--Boltzmann conductivity $\sigma_{DC}$ 
(intra-band conductivity) is evaluated
in the
relaxation time approximation.
With a realistic value of
scattering time, $\tau \sim 10^{-14}$\,s~\cite{Mayou93}, 
one
obtain $\sigma_{DC} \sim 300-1000\,({\rm \Omega cm})^{-1}$ for a
$\alpha$-Al--Mn model~\cite{Fujiwara93} and 1/1-Al--Fe--Cu
model~\cite{PRB94_AlCuFe}. 
These value correspond to the measured
values~\cite{Berger94} which are anomalously low for
metallic alloys.
For
decagonal approximant the anisotropy found experimentally in the
conductivity is also reproduced correctly~\cite{PRB94_AlCuCo}.

The semi-classical Block--Boltzmann description of transport gives
interesting results for the intra-band conductivity in crystalline
approximants, but it is insufficient to take into account most of
the aspect due to the special localization of electrons due to
the quasiperiodicity \cite{Roche97_JMP}.
Some specific transport
properties like the temperature dependence of the conductivity
(inverse Mathiessen rule, the defects influence,
the proximity of with a metal~/ insulator transition) requires to go
beyond a Block--Boltzmann analysis.
In fact, two different unconventional transport
mechanisms specific of these materials have been proposed
\cite{Mayou93,Fujiwara93,Roche97_JMP,Quasicrystals2000_Mayou}.
Transport could be dominated, for short relaxation time $\tau$ by
hopping between ``{\it critical} localized states'', whereas for
long time $\tau$ the regime could be dominated by non-ballistic
propagation of wave packets between two scattering events.

The experimental optical conductivity of quasicrystals
and approximants is also unusual.
The real part of the conductivity increases
linearly with the energy
at low energies (below 1 eV),
and there is no Drude
peak.
The absence of a Drude peak in approximants should be due to
the small DOS and the very low intra-band conductivity.
T. Fujiwara et al. \cite{Fujiwara93}
calculated the optical conductivity due to the
inter-band transition
from the LMTO band dispersion of $\alpha$-Al--Mn.
It reproduces the linearity and
the peak position observed.
More recently, D. Mayou~\cite{MayouPRL00}
derived a generalized
Drude formula for the optical conductivity of
quasicrystals.
It shows how a non-ballistic propagation
due to the quasiperiodicity can affect the optical
properties and explains the absence of a Drude peak,
the increase of conductivity with disorder,
and the inverse Mathiessen rule.

To conclude,
it appears that the degradation of the metallic character can
be obtained
either by a localization of states
or by a decrease of the DOS (semi-conducting state).
It seems that quasicrystals and approximants with large unit cell
combine both effects and this explains there
unusual transport properties.

\subsection{Cluster virtual bound states}

\begin{figure}[t!]
\begin{center}
\includegraphics[width=10cm]{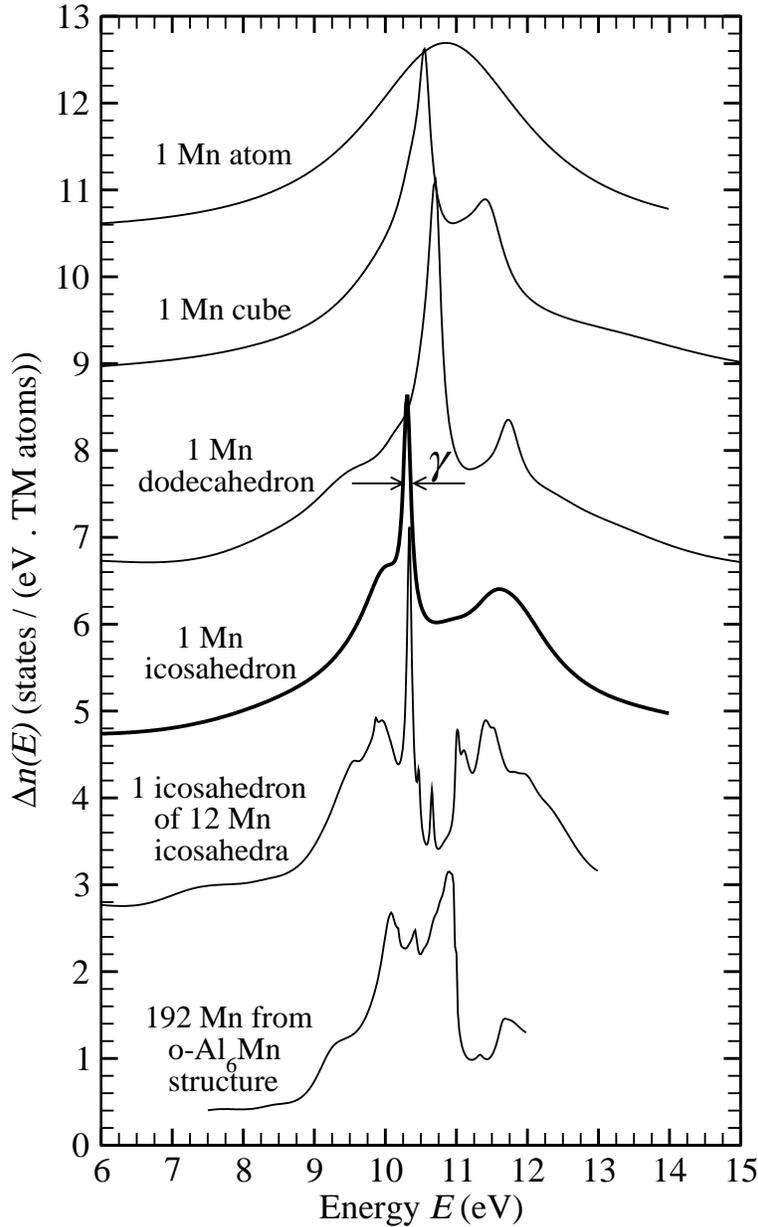}
\caption{Variation $\Delta n(E)$ of the DOS due to Mn clusters
in the Al matrix. In these clusters the first-neighbors Mn--Mn distances
are larger than 4.50\,$\rm \AA$.
Diameters $D$ of small clusters:
$D({\rm cube})= 3.90\,{\rm \AA}$,
$D({\rm icosahedron})=9.21 \,{\rm \AA}$
(actual value in $\alpha$-Al--Mn--Si,
$D({\rm dodecahedron})=12.33 \,{\rm \AA}$.
The icosahedron of 12 Mn icosahedra is obtained
after an inflation by a factor $\tau^2$ of an
initial Mn icosahedron, so the diameter of
the icosahedron of Mn icosahedra is
$\sim 33\,{\rm \AA}$ and it contains 144 Mn atoms. }
\label{DOS_cluster}
\end{center}
\end{figure}

The very low conductivity of quasicrystals and approximants
in spite of the non zero density of states at the Fermi level,
shows that electrons tend to be localized in a particular way.

The electronic structure of quasiperiodic lattices has
been studied theoretically within two different approaches.
The first consist in the analysis of the spectral properties
of quasiperiodic model Hamiltonian.
This approach shows the existence of a new kind of states
called critical states. Those states, which are neither
localized (like in doped-semiconductors) nor extended
(like in crystals), are characterized by the particular
localization due the long range quasiperiodic order.

The second approach to study the electronic properties
of quasicrystals consist in the study of realistic
approximants of quasicrystals.
By this way, we analyze the effect of the local atomic
order (chemical and topological local order) on the
quasicrystal properties. Band structure calculations
for approximants reveal that dispersion relations are flat,
corresponding to small velocities. Fine peaks in the density
of states (DOS) are associated with the flat bands.
Experimental results on transport properties show that
quasicrystals Al--Pd--Mn and Al--Cu--Fe and their periodic
approximants present very similar properties.
This suggests that the electronic transport in these alloys
is essentially determined by the local atomic order on the
length scale of the lattice parameter of the approximants,
i.e. $\rm 10 - 30 \AA$.

As for the local atomic order, one of the characteristics
of the quasicrystals and approximants, is the occurrence of
atomic clusters \cite{Gratias00}.
Nevertheless the clusters are not well defined because some
of them overlap each other, and they are a lot of glue atoms.
These remarks lead us to consider that clusters are not
isolated but they are embedded in metallic medium.
Our aim was to check whether the scattering of electrons
by cluster, on a scale of 10--30\,\AA,
can localize electrons.

Our model \cite{PRB97,RQ9_97}
is based on a standard description of inter-metallic alloys.
The central quantity is the transfer matrix ({\bf T} matrix)
of one cluster.
Considering the cluster embedded in a metallic medium,
we calculated the variation $\Delta n(E)$ of the DOS
due to the cluster (Lloyd formula).
For electrons, which have energy in the vicinity of
$E_{\rm F}$, transition atoms (such as Mn and Fe)
are strong scatterer whereas Al atoms are weak scatterer.
Then, following a classical approximation, we neglected the
potential of Al atoms.

In the figure \ref{DOS_cluster},
the variation $\Delta n(E)$ of the DOS due to different
clusters are shown. The Mn icosahedron is the actual Mn
icosahedron of the $\alpha$-Al--Mn--Si approximant.
As an example of a larger cluster, we consider one
icosahedron of Mn icosahedra, which might appeared in
the structural model for quasicrystals.

$\Delta n(E)$ of clusters exhibits strong deviations
from the Virtual Bound State corresponding to one Mn atom.
Indeed several peaks and shoulders appear.
The width of the most narrow peaks ($50 - 100$\,meV)
are comparable to the fine peaks of the calculated DOS
in the approximants.
Each peak corresponds to a resonance due to the scattering by the cluster.
This is associated to states \textit{``localized''}  by the cluster.
They are not eigenstate, they have finite lifetime
of the order of $\hbar/\gamma$, where $\gamma$ is
the width of the peak.
Therefore, the stronger the effect of the localization
by cluster is, the narrower is the peak.
The large lifetime is the proof of a localization,
but in the real space these states have an extension
on length scale of the cluster,
typically $\sim 9.21 \,{\rm \AA}$ for Mn icosahedron
that exist in the actual $\alpha$-Al--Mn--Si approximants.
As an example we have considered also
an icosahedron of 12 Mn icosahedra.
The diameter of each Mn icosahedron is  9.21\,$\rm \AA$ too, and the
diameter of the icosahedron of Mn icosahedra is $\sim 33$\,\AA.
The DOS of the large cluster contains new peaks with respect to
the simple Mn icosahedron (figure \ref{DOS_cluster}).
These are states localized on the length scale of about
33\,\AA. Therefore, in this large cluster,
some states are localized on the length scale of the
Mn icosahedron and other states are localized on the
length scale of the icosahedron of Mn icosahedra.

We named this new kind  of electronic states the
\textit{``cluster virtual bound states''}, by analogy with the
Virtual Bound States of
Friedel\,\cite{Friedel56}-Anderson\,\cite{Anderson61}
for a TM impurity.
Indeed,
the physical origin of these states can be understood as follows.
Consider incident electrons, with energy $E$ close to
$E_{\rm F}$,
scattered by the cluster.
In Al--Mn alloys
$E_{\rm F} \simeq E_d$,
where $E_d$ is the energy of the $d$ orbital. In this energy
range, the potential of the Mn atom is strong and the Mn atoms can
roughly be consider as hard spheres with radius of the order of the
$d$ orbital size ($\rm \sim 0.5\,\AA$).
An effect similar to that of the ``Faraday
cage'' can confine electrons in the cluster provided that their
wavelength $\lambda$ satisfies $\lambda \gtrsim l$,
where $l$ is the distance between two hard
spheres (TM--TM distances).
In the case of $\alpha$-Al--Mn--Si,
$l \simeq 3.6\,{\rm \AA}$ (if we assume a free
electron band and $E_{\rm F}= 10.33$\,eV)
and the distance $l$ is about 3.8\,$\rm \AA$.
Consequently, we expected to observe such a confinement.
This
effect is a multiple scattering effect, and it is not due to an
overlap between $d$ orbitals because Mn atoms are not
first-neighbor.

We have also  shown that these resonances are very sensitive to
the geometry of the cluster.
For instance, they disappear quickly
when the radius of the Mn icosahedron increases, or when the Mn
icosahedron is truncated \cite{RQ9_97}.

\subsection{Band-gap in some Al--TM alloys}

\begin{figure}[t]
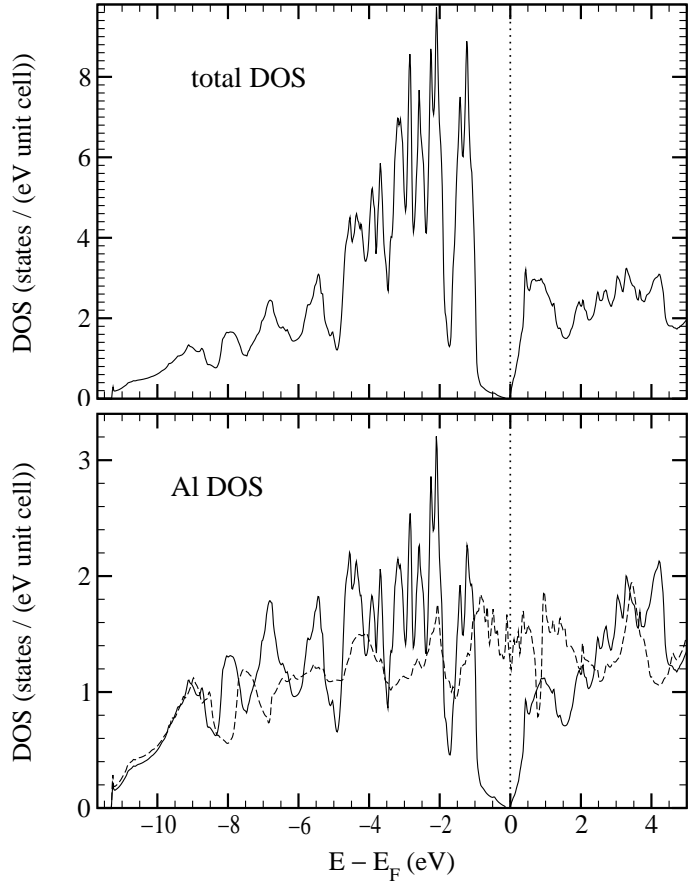

\begin{center}
\includegraphics[width=9cm]{c_al2ru_dost.eps}

\includegraphics[width=9cm]{c_al2ru_Al.eps}

\caption{LMTO total DOS and Al local DOS of
the orthorhombic $\rm Al_2Ru$.
Dashed line: calculation without $sp$-$d$ hybridization. \cite{Duc92}}
\label{Fig_LMTO_Al2Ru}
\end{center}
\end{figure}

While alloys composed of metallic constituents are expected to be metallic,
several Al--TM  phases are semi-conducting phases
with a band-gap smaller than traditional semiconductors (less than 1\,eV).
Experimental measurements of transport  properties and optical properties
indicate the presence of a small gap in
the DOS of the orthorhombic  $\rm Al_2Ru$
(C54 structure) \cite{Basov94_PRL}.
Its width is expected to be $\sim 0.17$\,eV.
This band-gap (or a very deep pseudogap) has been
confirmed theoretically from our
first-principles calculations \cite{Duc92}
and then other groups (see Refs. in \cite{PMS}).
All these works conclude that the band-gap is due the $sp$-$d$
hybridization,
but not
to charge transfers which are small \cite{Duc92}.
The DOS of $\rm Al_2Ru$ is shown on figure~\ref{Fig_LMTO_Al2Ru}.
A strong $p$(Al)-$d$(Ru) hybridization for electrons near $E_{\rm F}$
has also been
confirmed by photo-emission spectroscopy \cite{Fournee97}.

As shown recently by M. Kraj\v{c}\'{\i}
and J. Hafner  \cite{Krajci05_TMS}
from first-principles,
the semi-conducting gap
in $\rm Al_2TM$ DOS does
not disappear if TM sites are occupied by two different TM elements
($\rm TM_1$ and $\rm TM_2$), provided that the electron
per atom ratio is conserved.
These phases have hypothetical structures and thus this
does not prove that actual phases with the same composition
are semi-conducting phases.
From a detailed analysis of the ab initio calculations,
the authors
have shown an enhanced charge density halfway between
certain first-neighbor pairs of atoms, and
a bonding\,/\,anti-bonding splitting of the electronic states.
This suggests a dominantly covalent character of the bond
between atoms due to $sp$-$d$ hybridized orbitals.

A narrow gap may also be found in some hypothetical
Al(Si)--Mn phases.
Indeed, for particular positions of the Si atoms in
the $\alpha$-Al--Mn--Si phase,
a very narrow gap at $E_{\rm F}$ has been predicted very recently
by E.S. Zijlstra and S.K. Bose
\cite{Zijlstra03}.
We calculated \cite{PMS} DOS of $\delta$-$\rm Al_{11}Mn_4$
with the atomic structure proposed by Kontio {\it et al.}
(triclinic, $P\overline{1}$).
It exhibits a gap at energy close to $E_{\rm F}$.
According to recent structure refinement
this triclinic structure proposed
is not a good refinement of the crystallographic data.
But, it gives an interesting example for a possible
band-gap in Al(rich)--Mn materials.

\section{Conclusion}

Our theoretical studies on Al(rich)--transition-metal (TM) alloys lead to consider these
aluminides as $spd$ electron phases,
where a specific electronic structure governs the stability,
the electronic properties and the magnetism.
Schematically, the conduction states of these compounds
could be seen like $sp$ free states
(mainly Al states) scattered by the strong potential
of the TM atoms ($d$ orbitals).
The large value of the electronic density of the conduction electrons
($\sim 3$ conduction electrons per Al atoms)
creates strong Friedel oscillations of the charge density around each TM atoms.
Consequently, effective TM--TM interactions mediated by the $sp$
states are essential over medium-range distances (typically 10--20\,\AA).
In agreement with a Hume-Rothery minimization of the band energy,
these oscillating interactions leads to ``frustration'' mechanism which
favors complex atomic structures (including quasicrystals and
approximants phases).
Indeed, a specificity of these compounds is that
the stability (or not) of a TM atom on a given atomic site
does not depend only on the local environment, but it depends
also on TM--TM interactions over distances larger than
the first-neighbor distances.
This can explain preferred TM--TM spacings in Al(rich)--TM alloys and
the occurrence of atomic vacancies.
The occurrence of localized magnetic moments carried by the TM atoms
depends also on TM position via the
TM--TM interactions.
Our studies of the density of states gives also a simple explanation 
of the long standing
problem of the negative valence of TM atoms in these materials.
The strong scattering of the $sp$ states by the TM atoms could also
``localize'' conduction states on atomic clusters with diameter
of 10--30\,$\rm \AA$ and even more.
In some cases the system might go to a semi-conducting
regime with a gap in the density of states.
This gap is also due to the scattering of $sp$ states by $d$ orbitals.

\section{Acknowledgements}
Our work owes much to the discussions with
Prof. J. Friedel,
Prof. T. Fujiwara,
and Prof. D.G. Pettifor,
and to their works in the field of the electronic structure
of Hume--Rothery alloys.
We are also grateful to
many colleagues with whom we have collaborations:
M. Audier,
E. Belin--Ferr\'e,
C. Berger,
R. Bellissent,
A.M. Bratkovsky,
F. Cyrot--Lackmann,
F. Hippert,
J.P. Julien,
A. Pasturel,
J.J. Pr\'ejean,
L. Magaud,
S. Roche and
V. Simonet.
%
%
%
%

\bibliographystyle{plain}

\end{document}